\documentclass[twocolumn]{aastex63}

\usepackage{threeparttable, booktabs}

\newcommand{\feh}{[Fe/H]}
\newcommand{\alphafe}{[$\alpha$/Fe]}
\newcommand{\fehavg}{$\langle$[Fe/H]$\rangle$}
\newcommand{\alphafeavg}{$\langle$[$\alpha$/Fe]$\rangle$}
\newcommand{\kms}{km s$^{-1}$}
\newcommand{\teffphot}{$T_{\rm{eff,phot}}$}
\newcommand{\loggphot}{$\log$ $g$}
\newcommand{\fehphot}{[Fe/H]$_{\rm{phot}}$}
\newcommand{\teff}{$T_{\rm{eff}}$}
\newcommand{\logg}{$\log$ $g$}

\newcommand{\sgrcorestreamalphadiff}{$0.06-0.15$ dex}
\newcommand{\sgrcorestreamalphagrad}{$(0.1-1.2) \times 10^{-3}$ dex deg$^{-1}$}
\newcommand{\lgdwarfmaxfehgrad}{$-0.4$ dex per half-light radius}
\newcommand{\mwdiskfehgrad}{$-0.06$ dex kpc$^{-1}$}
\newcommand{\allfehgrad}{$-(0.06-0.08)$ dex kpc$^{-1}$} 

\newcommand{\sgrfehgradage}{$-0.12$ $\pm$ 0.03 dex Gyr$^{-1}$}
\newcommand{\sgrstreamintfehgrad}{$-(1.2-1.4) \times 10^{-3}$ dex deg$^{-1}$}
\newcommand{\sgrprogfehgrad}{$-0.2$ dex kpc$^{-1}$}
\newcommand{\sgrcorestreamfehdiff}{$0.4-0.6$}
\newcommand{\fehphotspecdiffthirtythreekpcGSSfield}{0.60}
\newcommand{\seomfehphotspecdiffthirtythreekpcGSSfield}{0.11}
\newcommand{\fehsynthcorevsandbias}{0.26}
\newcommand{\fehsynthcorevsathirteenbias}{0.57}
\newcommand{\mfiftyfehphotdifflo}{0.60}
\newcommand{\mfiftyfehphotdiffhi}{0.90}
\newcommand{\kalirairajafehphotdiff}{0.09}
\newcommand{\fehsynthcorevsathirteen}{1.62}
\newcommand{\fehsynthcorevsathirteenerr}{0.48}
\newcommand{\fehsynthcorevsand}{0.63}
\newcommand{\fehsynthcorevsanderr}{0.10}
\newcommand{\fehphotcorevsathirteen}{0.56}
\newcommand{\fehphotcorevsathirteenerr}{0.15}
\newcommand{\fehphotcorevsand}{$-$0.10}
\newcommand{\fehphotcorevsanderr}{0.05}
\newcommand{\gilcoreandfehphotdiffmed}{0.10} 
\newcommand{\gilcoreathirteenfehphotdiffavg}{0.53} 
\newcommand{\gilcoreathirteenfehphotdiffavgerr}{0.13}
\newcommand{\ngssouterphot}{339}
\newcommand{\ngssphot}{270}

\newcommand{\ntotgss}{62}

\newcommand{\ntotouthalo}{203}
\newcommand{\ntotpub}{200}
\newcommand{\gssfehgradient}{$-$0.018}
\newcommand{\gssfehgradienterr}{0.003}
\newcommand{\dm}{24.63}
\newcommand{\ddm}{0.2}
\newcommand{\vsysmed}{1.49}
\newcommand{\vsyslow}{5.6}
\newcommand{\fehsyslow}{0.130}
\newcommand{\alphafesyslow}{0.107}
\newcommand{\fehsysmed}{0.101}
\newcommand{\alphafesysmed}{0.084}

\newcommand{\gssdistS}{23}
\newcommand{\gssdistathree}{38}
\newcommand{\deltateffS}{1.33}
\newcommand{\deltatefferrS}{0.79}
\newcommand{\deltaloggS}{$-$0.02}
\newcommand{\deltaloggerrS}{0.04}
\newcommand{\deltateffathree}{0.89}
\newcommand{\deltatefferrathree}{0.99}
\newcommand{\deltaloggathree}{$-$0.04}
\newcommand{\deltaloggerrathree}{0.04}
\newcommand{\deltadmS}{0.06}
\newcommand{\deltadmathree}{0.10}
\newcommand{\deltafehathree}{$-$0.01}
\newcommand{\deltafeherrathree}{0.03}
\newcommand{\deltaalphaathree}{0.01}
\newcommand{\deltaalphaerrathree}{0.07}
\newcommand{\deltateffspecathree}{0.98}
\newcommand{\deltateffspecerrathree}{0.40}
\newcommand{\deltafehS}{0.0}
\newcommand{\deltafeherrS}{0.12}
\newcommand{\deltaalphaS}{$-$0.01}
\newcommand{\deltaalphaerrS}{0.27}
\newcommand{\deltateffspecS}{6.36}
\newcommand{\deltateffspecerrS}{2.51}

\received{15 Mar 2021}
\revised{5 May 2021}
\accepted{5 May 2021}

\shorttitle{GSS Abundance Gradients}
\shortauthors{Escala et al.}

\graphicspath{{./}{figures/}}

\begin{document}

\title{Elemental Abundances in M31: Gradients in the Giant Stellar Stream\footnote{The data presented herein were obtained at the W. M. Keck Observatory, which is operated as a scientific partnership among the California Institute of Technology, the University of California and the National Aeronautics and Space Administration. The Observatory was made possible by the generous financial support of the W. M. Keck Foundation.}}

\correspondingauthor{I. Escala}
\email{iescala@carnegiescience.edu}

\author[0000-0002-9933-9551]{Ivanna Escala}
\altaffiliation{Carnegie-Princeton Fellow}
\affiliation{The Observatories of the Carnegie Institution for Science, 813 Santa Barbara St, Pasadena, CA 91101, USA}
\affiliation{Department of Astrophysical Sciences, Princeton University, 4 Ivy Lane, Princeton, NJ 08544, USA}

\author[0000-0003-0394-8377]{Karoline M. Gilbert}
\affiliation{Space Telescope Science Institute, 3700 San Martin Drive, Baltimore, MD 21218, USA}
\affiliation{Department of Physics \& Astronomy, Bloomberg Center for Physics and Astronomy, John Hopkins University, 3400 N. Charles St, Baltimore, MD 21218, USA}

\author[0000-0002-3233-3032]{Jennifer Wojno}
\affiliation{Department of Physics \& Astronomy, Bloomberg Center for Physics and Astronomy, John Hopkins University, 3400 N. Charles St, Baltimore, MD 21218, USA}

\author[0000-0001-6196-5162]{Evan N. Kirby}
\affiliation{Department of Astronomy, California Institute of Technology, 1200 E California Blvd, Pasadena, CA 91125, USA}


\author[0000-0001-8867-4234]{Puragra Guhathakurta}
\affiliation{UCO/Lick Observatory, Department of Astronomy \& Astrophysics, University of California Santa Cruz, 1156 High Street, Santa Cruz, CA 95064, USA}

\begin{abstract}

We analyze existing measurements of \feh\ and \alphafe\ for individual red giant branch (RGB) stars in the Giant Stellar Stream (GSS) of M31 to determine whether spatial abundance gradients are present. These measurements were obtained from low-- ($R \sim 3000$) and moderate-- ($R \sim 6000$) resolution Keck/DEIMOS spectroscopy using spectral synthesis techniques as part of the Elemental Abundances in M31 survey. From a sample of \ntotgss\ RGB stars spanning the GSS at 17, 22, and 33 projected kpc, we measure a \feh\ gradient of \gssfehgradient\ $\pm$ \gssfehgradienterr\ dex kpc$^{-1}$ and negligible \alphafe\ gradient with M31-centric radius. 
We investigate GSS abundance patterns in the outer halo using additional \feh\ and \alphafe\ measurements for 6 RGB stars located along the stream at 45 and 58 projected kpc. 
These abundances provide tentative evidence that the trends in \feh\ and \alphafe\ beyond 40 kpc in the GSS are consistent with those within 33 kpc.
We also compare the GSS abundances to 65 RGB stars located along the possibly related Southeast (SE) shelf substructure at 12 and 18 projected kpc. The abundances of the GSS and SE shelf are consistent, supporting a common origin hypothesis, although this interpretation may be complicated by the presence of  \feh\ gradients in the GSS\@. We discuss the abundance patterns in the context of photometric studies from the literature and explore implications for the properties of the GSS progenitor, suggesting that the high $\langle$\alphafe$\rangle$ of the GSS (+0.40 $\pm$ 0.05 dex) favors a major merger scenario for its formation.

\end{abstract}


\section{Introduction} \label{sec:intro}

Stellar streams originate from the ongoing tidal disruption of accreted galaxies and globular clusters, providing an instantaneous view of the hierarchical formation of the host galaxy (e.g., \citealt{FreemanBland-Hawthorn2002,BullockJohnston2005,Helmi2020}). In the Milky Way (MW), the discovery of the Sagittarius stream \citep{Ibata2001b} 
provided an early indication of the importance of mergers in Galactic formation history. The contemporaneous discovery of M31's Giant Stellar Stream (GSS;  \citealt{Ibata2001})
 further indicated that stellar streams are a common feature of galaxies beyond the MW, and that mergers have also played a significant role in M31's evolution.
 
 The GSS is a conspicuous tidal structure in M31's southeastern quadrant that spans at least 6 degrees ($\sim$80 projected kpc) on the sky and $\gtrsim$100 kpc in line-of-sight distance over its extent \citep{McConnachie2003,Conn2016}. The stream appears to be characterized by a metal-rich, high surface brightness core ($\Sigma_V \sim 30$ mag arcsec$^{-2}$; \citealt{Ibata2001}) and an asymmetric envelope that has both lower metallicity and surface brightness \citep{Ibata2007}. In comparison to the phase-mixed component of M31's stellar halo, photometric and spectroscopic studies of the GSS's resolved stellar populations have revealed that it is more metal-rich, kinematically colder, and possesses more dominant intermediate-age stellar populations (e.g., \citealt{Guhathakurta2006,Kalirai2006d,Brown2006,Ibata2007,Gilbert2009,Tanaka2010,Ibata2014}). Based on these properties, the GSS was inferred to originate from the recent ($\lesssim1$ Gyr) disruption of a distinct satellite progenitor on a highly radial orbit with a lower stellar mass limit of $\sim10^8 M_\odot$ \citep{Ibata2004,Font2006,Fardal2006}. 
 
However, the nature of the GSS accretion event is likely more complex than initially surmised.  Spectroscopic surveys of M31's stellar halo have uncovered a number of faint kinematical features that are tidal debris possibly related to the GSS. \citet{Kalirai2006d} first detected a second kinematically cold component (KCC) in a field probing the GSS at 20 projected kpc that was not a prediction of concurrent dynamical models \citep{Ibata2004,Font2006,Fardal2006} despite the similarity of its photometric metallicity to the primary GSS substructure. \citet{Gilbert2009} later traced the KCC inward to 17 projected kpc, showing that the feature was consistently separated from the GSS by $\sim$100 km s$^{-1}$ in line-of-sight velocity over its spanned radial range, thus providing compelling evidence in favor of a direct physical connection between the GSS and KCC. 

Following the discovery of the KCC, \citet{Gilbert2007} kinematically detected a faint substructure component located
 $\sim$11--18 projected kpc along M31's southeastern minor axis. Unlike in the case of the KCC, this feature matched predictions from models of the GSS accretion event; specifically, for the Southeast (SE) shelf generated by the fourth pericentric passage of the GSS progenitor \citep{Fardal2006,Fardal2007}. The similarity of the photometric metallicity and age distributions of stellar populations in the SE shelf and GSS \citep{Brown2003,Brown2006,Gilbert2007} further bolstered the hypothesis that the SE shelf and GSS were tidal debris from the same event. The prediction of the SE shelf illustrates that minor merger models for the formation of the GSS ($M_\star \sim (1-5) \times 10^{9} M_\odot$; \citealt{Fardal2006,Fardal2007,Fardal2008,Fardal2013,MoriRich2008,Sadoun2014,Kirihara2014,Kirihara2017, Miki2016}) can successfully reproduce the broad morphological and kinematical features of the stream while accounting for diffuse shell-like features such as the Northeast \citep{Ferguson2002,Ferguson2005} and West (W; \citealt{Fardal2007}) shelves as part of the forward continuation of the stream. In further support of this hypothesis, \citet{Fardal2012} showed that the kinematics of the W shelf were strikingly similar to predictions for the feature, and that the shelf's metallicity was consistent with that of the GSS.
 
Nevertheless, minor merger models for the GSS's formation are unable to simultaneously provide a concise explanation for the origin of the KCC. \citet{Gilbert2019} speculated that an asymmetric extension of the W shelf toward M31's SE quadrant could potentially account for the KCC within this framework, although multiple superposed loops of the GSS also provide a feasible explanation for the KCC in a major merger scenario ($M_\star \sim 10^{10} M_\odot$; \citealt{Hammer2010,Hammer2018,DSouzaBell2018}). The formation of the GSS via a major merger had not been explored earlier in order to preserve the integrity of M31's disk (e.g., \citealt{MoriRich2008}), though simulations have since demonstrated that gas-rich mergers can enable disk survival (e.g., \citealt{Hopkins2009}). Without disk intactness as a constraining factor, the GSS and its associated shells can be reproduced by merger ratios varying from 300:1 to 2:1 \citep{Hammer2018}, casting uncertainty on whether a major ($<$10:1) or minor ($\geq$10:1) merger is responsible for the stream.
 
 Chemical abundance measurements (\feh\ and \alphafe) of individual red giant branch (RGB) stars in the GSS have the potential to elucidate the properties of the progenitor by breaking the degeneracy between formation models. Simulations of MW-mass galaxies have shown that the mass and accretion time distributions of external progenitors can imprint strong chemical signatures in a galaxy's accreted stellar populations in terms of Fe and $\alpha$-elements (O, Ne, Mg, Si, S, Ar, Ca, and Ti), respectively (e.g., \citealt{Robertson2005,Font2006abund,Johnston2008}). Using an extrapolation of the stellar mass metallicity relation for Local Group dwarf galaxies \citep{Kirby2013}, \citet{Gilbert2019} estimated a stellar mass for the progenitor of $(1-5)\times10^9 M_\odot$ based on the first spectral synthesis based \feh\ measurements from a field located at 17 projected kpc in the GSS. \citeauthor{Gilbert2019} also found that the GSS has a high average $\alpha$-enhancement ($\sim0.4$ dex), indicating that its progenitor formed stars with high efficiency. \citet{Escala2020a} later confirmed that the chemical abundance patterns found by \citet{Gilbert2019} extended to a GSS field at 22 projected kpc. Although the stellar mass predicted by iron abundance in the GSS is consistent with minor merger models, this cannot be interpreted as direct evidence in favor of \deleted{a} such a scenario if the progenitor had a metallicity gradient.
 
 Indeed, massive satellite galaxies in the Local Group such the \added{Large and Small} Magellanic Clouds \added{(LMC and SMC)}
 , M33, 
 and Sagittarius \added{(Sgr)}
 are known to possess negative radial metallicity gradients in their RGB populations. \added{For the LMC, SMC, and M33, radial metallicity gradients of \allfehgrad\ have been detected out to several disk scale lengths (LMC: \citealt{Choudhury2016}; SMC: \citealt{Dobbie2014b,Parisi2016,Choudhury2018}; M33: \citealt{Kim2002,Tiede2004,Barker2007})
 that are similar to the gradient in the MW's disk (\mwdiskfehgrad; e.g., \citealt{Cheng2012a,Hayden2014}). In addition, metallicity differences of \sgrcorestreamfehdiff\ have been observed between the Sgr core and Sgr streams \citep{Chou2007,Monaco2007,Keller2010,Hayes2020}, which translate to an intrinsic 
 gradient of about \sgrprogfehgrad\ in the Sgr progenitor based on dynamical modeling \citep{LawMajewski2010}.\footnote{Possible explanations for this steep gradient include a massive, disky Sgr progenitor similar to the Large Magellanic Cloud, or more likely, the lack of self-consistent modeling of continuing star formation in the Sgr core as the progenitor tidally disrupts over several Gyr \citep{Hayes2020}.} Although only weak internal gradients are measured along the Sgr streams (\sgrstreamintfehgrad, but consistent with zero within the uncertainties; \citealt{Hayes2020}), combining these measurements with modeling provides evidence for a gradient with dynamical age (i.e., initial orbital radius; \sgrfehgradage) in the Sgr progenitor.
 In comparison, smaller Local Group dwarf galaxies (M$_\star \lesssim 10^{8.5} M_\odot$) have diverse gradients, ranging from flat to as steep as \lgdwarfmaxfehgrad\  (e.g., \citealt{Kirby2011,Kirby2017,Leaman2013,Vargas2014a,Ho2015,Kacharov2017}), with no clear relationship between the magnitude of a gradient and luminosity, host distance, or morphology (\citealt{Ho2015}, c.f. \citealt{Leaman2013}), although there may be a trend with median stellar age \citep{Mercado2021}.}
 
 In accordance with expectations based on the \added{GSS} progenitor's inferred mass, an early photometric survey of stellar populations along the line-of-sight to the GSS \citep{Ferguson2002} noted the presence of color variations over the stream, which were attributed to metallicity variations. Subsequently, \citet{Ibata2007} inspected such variations between the high surface brightness core of the GSS and its extended envelope, noting that the latter had lower average photometric metallicity. \citet{Gilbert2009} provided additional support for this dichotomy by measuring photometric metallicities of spectroscopically confirmed RGB stars in the core and envelope of the GSS. More recently, photometric studies have embarked on increasingly detailed explorations of GSS metallicity variations as a function of two-dimensional position on the sky \citep{Conn2016,Cohen2018}.
 
 Thus, detailed observations of abundance gradients in the GSS are necessary to map on-sky variations to the initial abundance properties of the progenitor. In this work, we present a comprehensive analysis of spatial \feh\ and \alphafe\ gradients in the GSS and likely associated substructures using spectral synthesis based abundance measurements from the Elemental Abundances in M31 survey \citep{Escala2019,Escala2020a,Escala2020b,Gilbert2019,Gilbert2020,Kirby2020,Wojno2020} with the aim of providing further constraints for GSS formation models. In Section~\ref{sec:data}, we provide an overview of the spectroscopic data and chemical abundance measurements, which we use to investigate the GSS's abundance properties between 17--58 kpc in Section~\ref{sec:grad}. We conclude by discussing our results in the context of both the observational and theoretical literature in Section~\ref{sec:discuss} before summarizing in Section~\ref{sec:summary}.

\section{Data} \label{sec:data}

\begin{figure*}
    \centering
    \includegraphics[width=\textwidth]{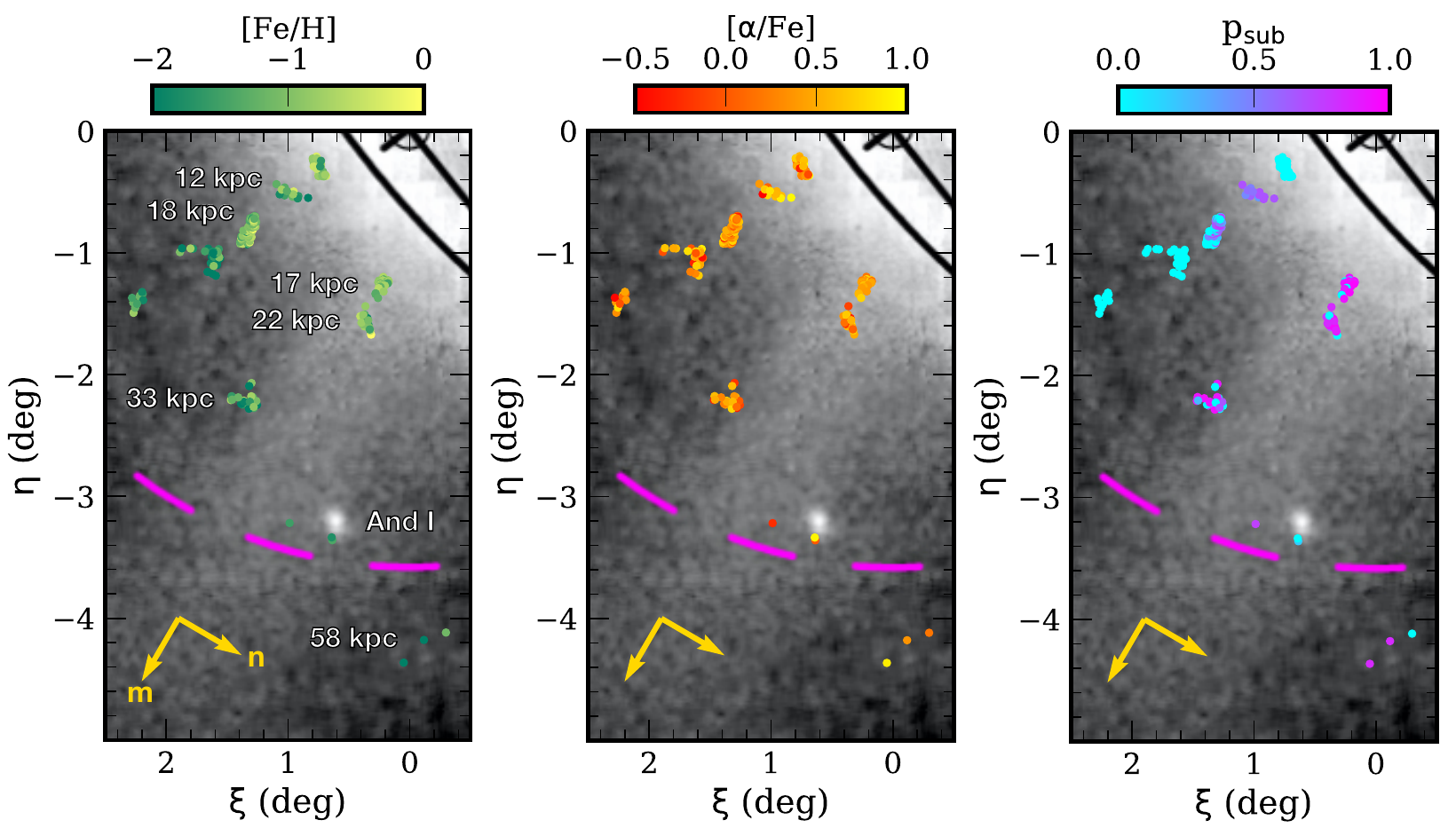}
    \caption{ The spatial distribution of \ntotouthalo\ RGB stars in M31's southeast quadrant, given in M31-centric coordinates, with measurements of [Fe/H] and [$\alpha$/Fe] from the Elemental Abundances in M31 survey (\citealt{Gilbert2019,Gilbert2020,Escala2020a,Escala2020b}; J.~Wojno et al. in preparation). The spectroscopic fields utilized in this work (Table~\ref{tab:fields}) are labeled according to their projected M31-centric distance, excepting And I at 45 projected kpc (left panel). The RGB stars are overlaid on the PAndAS star count map \citep{McConnachie2018}. In each panel, stars are color coded by (left) \feh, (middle) \alphafe, and (right) probability of belonging to kinematically cold substructure. The thick, solid black lines represent the edge of M31's classical disk ($i = 77^{\circ}$, $r = 17$ kpc) and the orientation of its minor axis. The dashed magenta lines delineate 50 projected kpc.
    The gold vectors represent GSS-aligned coordinate axes \citep{Fardal2006}.}
    \label{fig:gss_2d}
\end{figure*}

We utilized existing measurements of \feh\ and \alphafe\ for individual red giant branch (RGB) stars in M31's stellar halo obtained from low-- ($R \sim 3000$) and moderate-- ($R \sim 6000$) resolution Keck/DEIMOS spectroscopy as part of the Elemental Abundances in M31 survey \citep{Gilbert2019,Gilbert2020,Escala2019,Escala2020a,Escala2020b}. In total, \ntotpub\ RGB stars in our sample have published measurements of \feh\ and \alphafe\ in the southeastern quadrant of M31's stellar halo. We also include unpublished measurements (J.~Wojno et al., in preparation) for 3 M31 RGB stars in a spectroscopic field overlapping with the GSS envelope at 58 projected kpc. Figure~\ref{fig:gss_2d} illustrates the spatial distribution of these stars compared to the star count map from the Pan-Andromeda Archaeological Survey (PAndAS; \citealt{McConnachie2018}), while providing a sense of the variation in \feh\ and \alphafe\ over the probed region. 

In contrast to previous work by \citet{Escala2020b} using a nearly identical sample, we focused our analysis on M31 RGB stars with a high probability of belonging to kinematically identifiable substructure based on their heliocentric radial velocities ($p_{\rm sub}$; right panel of Figure~\ref{fig:gss_2d}; \S~\ref{sec:vel}). The majority of these stars are located in spectroscopic fields along the GSS at 17, 22, and 33 projected kpc from the center of M31, with a few stars located in the outer halo at 45 and 58 projected kpc. Additional stars with nonzero substructure probability, which are likely associated with the Southeast shelf substructure \citep{Gilbert2007,Fardal2007,Escala2020a} are located in fields at 12 and 18 kpc along M31's minor axis. We provide a brief summary of the spectroscopic observations and abundance measurements below, and refer the reader to \citet{Gilbert2019,Gilbert2020} and \citet{Escala2019,Escala2020a,Escala2020b} for further details.

\subsection{Spectroscopy}
\label{sec:spectra}

\begin{table*}
\begin{threeparttable}
\caption{Properties of Substructure in Spectroscopic Fields}
\label{tab:fields}
\begin{tabular*}{\textwidth}{l @{\extracolsep{\fill}} cccccccccc}
\hline
\hline
Field &  \multicolumn{1}{p{1.3cm}}{\centering $r_{\rm proj}$\\(kpc)} & Comp. & \multicolumn{1}{p{1.3cm}}{\centering $\mu$\\(\kms)}
& \multicolumn{1}{p{1.3cm}}{\centering $\sigma$\\(\kms)}
& $f$ & $\langle$[Fe/H]$\rangle$ &  $\langle$[$\alpha$/Fe]$\rangle$ & N$_{\rm M31}$ & N$_{\rm [\alpha/Fe]}$ & Ref.\\
\hline

f207 & 17 & GSS & $-$529.4 & 24.5 & 0.33 & $-$0.87$^{+0.09}_{-0.10}$ & +0.44$^{+0.04}_{-0.05}$ &  108 & 21 & 1,2\\
& & KCC & $-$427.3 & 21.0 & 0.32 & $-$0.79 $\pm$ 0.07 & +0.54 $\pm$ 0.06 & ... & ... & ...\\
S & 22 & GSS & $-$489.0 & 26.1 & 0.49 & $-$1.02$^{+0.15}_{-0.14}$ & +0.38$^{+0.17}_{-0.19}$ & 87 & 20 & 3\\
& & KCC & $-$371.6 & 17.6 & 0.22 & $-$0.71 $\pm$ 0.11 & +0.35$^{+0.08}_{-0.09}$ &  ... & ... & ...\\
a3 & 33 & GSS &  $-$444.6 & 15.7 & 0.56 &  $-$1.11$^{+0.12}_{-0.13}$ & +0.34$^{+0.08}_{-0.09}$ & 75 & 21 & 1,4\\
And I & 45 & GSS & $-$383.3 & 32.4 & 0.56 & $-$1.49$^{+0.10}_{-0.02}$ & $-$0.19$^{+0.21}_{-0.05}$ & 38 & 3 & 1,5\\
a13 & 58 & GSS &  $-$301.6 & 29.2 & 0.62 & $-$2.50$^{+0.40}_{-0.54}$ & +0.71$^{+0.19}_{-0.34}$ & 31 & 3 & 1,6,7\\
H & 12 & SE Shelf &  $-$295.4 & 65.8 & 0.56 & $-$1.30$^{+0.13}_{-0.12}$ & +0.53$^{+0.08}_{-0.10}$ & 104 & 16 & 3\\
f123 & 18 & SE Shelf & $-$279.9 & 11.0 & 0.32 & $-$0.71 $\pm$ 0.07 & +0.41$^{+0.04}_{-0.05}$ & 74 & 49 & 1,4\\

\hline
\end{tabular*}
\begin{tablenotes}
\item Note. \textemdash\ The columns of the table refer to spectroscopic field name, projected radius from M31's galactic center, substructure component, the median heliocentric velocity ($\mu$), dispersion ($\sigma$), and fractional contribution ($f$) of the velocity model for a given component (\S~\ref{sec:vel}), average \feh\ and \alphafe\ for a given component (\S~\ref{sec:abund_measure}), weighted by the inverse variance of the measurement uncertainty and substructure probability, number of spectroscopically confirmed RGB members (excluding dwarf galaxy members for And I; \S~\ref{sec:vel}), and number of RGB stars with successful \feh\ and \alphafe\ measurements. References: (1) \citet{Gilbert2018}, (2) \citet{Gilbert2019}, (3) \citet{Escala2020a}, (4) \citet{Escala2020b}, (5) \citet{Gilbert2020}, (6) J.~Wojno et al. in preparation, (7) this work.
\end{tablenotes}
\end{threeparttable}
\end{table*}


All spectroscopic fields, except a13 and And I\footnote{The field And I is based on a mixture of both shallow and deep spectroscopic data from the SPLASH survey \citep{Gilbert2009} and the Elemental Abundances in M31 survey \citep{Kirby2020}.}, were observed for a minimum of 5 hr with the 600 line mm$^{-1}$ or 1200 line mm$^{-1}$ grating for the case of low-- and moderate-- resolution spectroscopy, respectively. These configurations result in spectra with a FWHM spectral resolution of 2.8 \AA\ ($R\sim3000$) and 1.2 \AA\ ($R\sim6000$). Additionally, each deep (5+ hr) field was designed from previous shallow ($\sim$1 hr) DEIMOS observations from the Spectroscopic and Photometric Landscape of Andromeda's Stellar Halo (SPLASH) survey \citep{Guhathakurta2006,Kalirai2006d,Gilbert2007,Gilbert2009} in order to maximize the yield of spectroscopically confirmed M31 RGB stars. Data for fields a13 and And I were obtained as part of the SPLASH survey, where a handful of stars have spectra with fortuitously high signal-to-noise ratios such that measuring abundances is feasible. The spectra were reduced using a modified version of the spec2d pipeline (\citealt{Cooper2012,Newman2013} for the original pipeline; \citealt{SimonGeha2007,Kirby2020} for modifications specific to stellar point sources). Table~\ref{tab:fields} provides a summary of the properties for each spectroscopic field containing kinematically identifiable substructure.

\subsection{Radial Velocity Measurements and Membership Determination}
\label{sec:vel}

Radial velocities were measured via cross-correlation with templates using the procedures described in \citet{SimonGeha2007} and \citet{Kirby2015}. The statistical uncertainty is calculated from Monte Carlo trials in which a given observed spectrum is perturbed according to its standard error, whereas the systematic uncertainty is calculated via repeat velocity measurements of the same stars. We adopted a systematic velocity term of \vsysmed\ \kms\ for 1200 line mm$^{-1}$ grating spectra \citep{Kirby2015} and \vsyslow\ \kms\ for 600 line mm$^{-1}$ grating spectra \citep{Collins2011}. 

Using a combination of heliocentric radial velocity, color-magnitude diagram position, Na I $\lambda\lambda$8190 equivalent widths, and photometric and calcium-triplet based metallicity estimates, we assigned a probability of belonging to M31 to each star with a successful velocity measurement. We utilized the Bayesian inference method of \citet{Escala2020b} to determine membership for all stars except those in the 17, 45, and 58 kpc fields, for which we used the maximum likelihood based technique of \citet{Gilbert2006}. For the 45 kpc field, we additionally required that M31 RGB candidates had \fehphot\ $>$ $-0.95$ in order to separate M31 halo stars from stars belonging to the And I dwarf spheroidal galaxy. \citet{Gilbert2009,Gilbert2020} demonstrated that this additional photometric metallicity criterion clearly demarcates the two populations, regardless of velocity or apparent proximity to the dwarf galaxy.

As discussed by \citet{Escala2020b}, the Bayesian inference and maximum likelihood based methods of M31 membership determination produce generally consistent results, where \citeauthor{Escala2020b}'s classification of stars as M31 members is slightly more conservative. In general, we consider stars to be M31 RGB stars if they are more likely to belong to M31 than the MW foreground. The membership criterion for the 45 and 58 kpc fields is more stringent, requiring that stars are at least three times more likely to belong to M31 than the MW, owing to the increased likelihood of contamination by MW foreground dwarfs in M31's sparse outer halo. 

\begin{figure*}
    \centering
    \includegraphics[width=\textwidth]{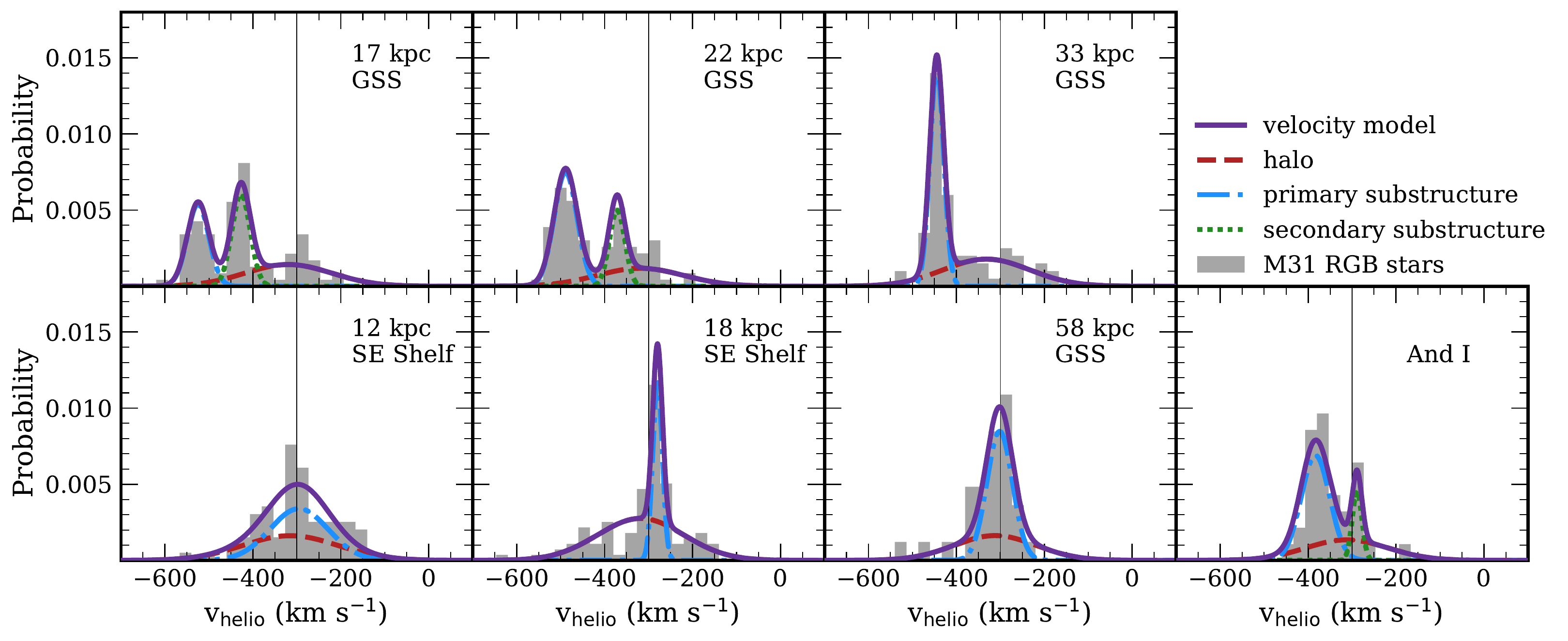}
    \caption{Heliocentric radial velocity distributions (\S~\ref{sec:vel}) for M31 RGB stars (grey histograms) in spectroscopic fields with detected kinematic substructure (Table~\ref{tab:fields}; \citealt{Gilbert2019,Escala2020a,Escala2020b}). We show the adopted velocity model for each field (purple solid lines; \citealt{Gilbert2018,Escala2020a}), including kinematically hot halo components (dashed red lines), and cold components (dashdotted blue lines and dotted green lines) corresponding to primary and secondary substructures. The substructure components present in these fields are the GSS, KCC, and SE shelf. The And I field at 45 projected kpc contains a dwarf galaxy, but also overlaps with the GSS (Table~\ref{tab:fields}). RGB stars that are likely And I members are excluded from the field's velocity distribution.}
    \label{fig:vels}
\end{figure*}

Figure~\ref{fig:vels} shows the heliocentric radial velocity distribution of RGB stars in each spectroscopic field containing substructure (Table~\ref{tab:fields}), where stars with velocities consistent with that of kinematically cold components have a higher probability of belonging to substructure ($p_{\rm sub}$; \citealt{Gilbert2019,Escala2020a,Escala2020b}). Figure~\ref{fig:vels} also shows Gaussian mixture models of the velocity distribution for each field, where each model contains both halo and substructure components. We adopted the 50$^{\rm th}$ percentile values of the marginalized posterior probability distributions from \citet{Escala2020a} to model the substructure components in the 12 kpc and 22 kpc fields, whereas all other component models (including halo components) are from \citet{Gilbert2018}. The substructure probability for a star with a given radial velocity is thus the odds ratio of the Bayes factor under the assumption of the substructure versus halo models.

\subsection{Chemical Abundance Measurements} \label{sec:abund_measure}


Chemical abundance (\feh\ and \alphafe) and stellar parameter measurements (\teff) were obtained from spectral synthesis of low- and medium-resolution stellar spectroscopy for individual RGB stars in each field. In summary, each observed spectrum is compared to a grid of synthetic spectra using Levenberg-Marquardt minimization to identify the best-fit stellar parameters and abundances. Throughout this procedure, the spectroscopic effective temperature (\teff) is loosely constrained by photometry, whereas the surface gravity (\logg) is fixed to its photometric value, assuming a distance modulus of $(m - M)$ = \dm\ $\pm$ \ddm\ \citep{Clementini2011} for M31. 
Measurements of \feh\ and \alphafe\ obtained for identical stars from low- and medium-resolution spectra (\S~\ref{sec:spectra}) are generally consistent within the uncertainties \citep{Escala2020a}.  Systematic uncertainties on the abundance measurements are added in quadrature to the random component of the uncertainty from the fitting procedure. We adopted systematic error terms of \fehsyslow\ (\fehsysmed) and \alphafesyslow\ (\alphafesysmed) for \feh\ and \alphafe\ measurements, respectively, obtained from 600 (1200) line mm$^{-1}$ spectra \citep{Escala2020a,Gilbert2019}. We refer the reader to \citet{Escala2019,Escala2020a} and \citet{Kirby2008,Kirby2009} for detailed descriptions of the low- and medium-resolution spectral synthesis techniques.

\begin{figure}
    \centering
    \includegraphics[width=\columnwidth]{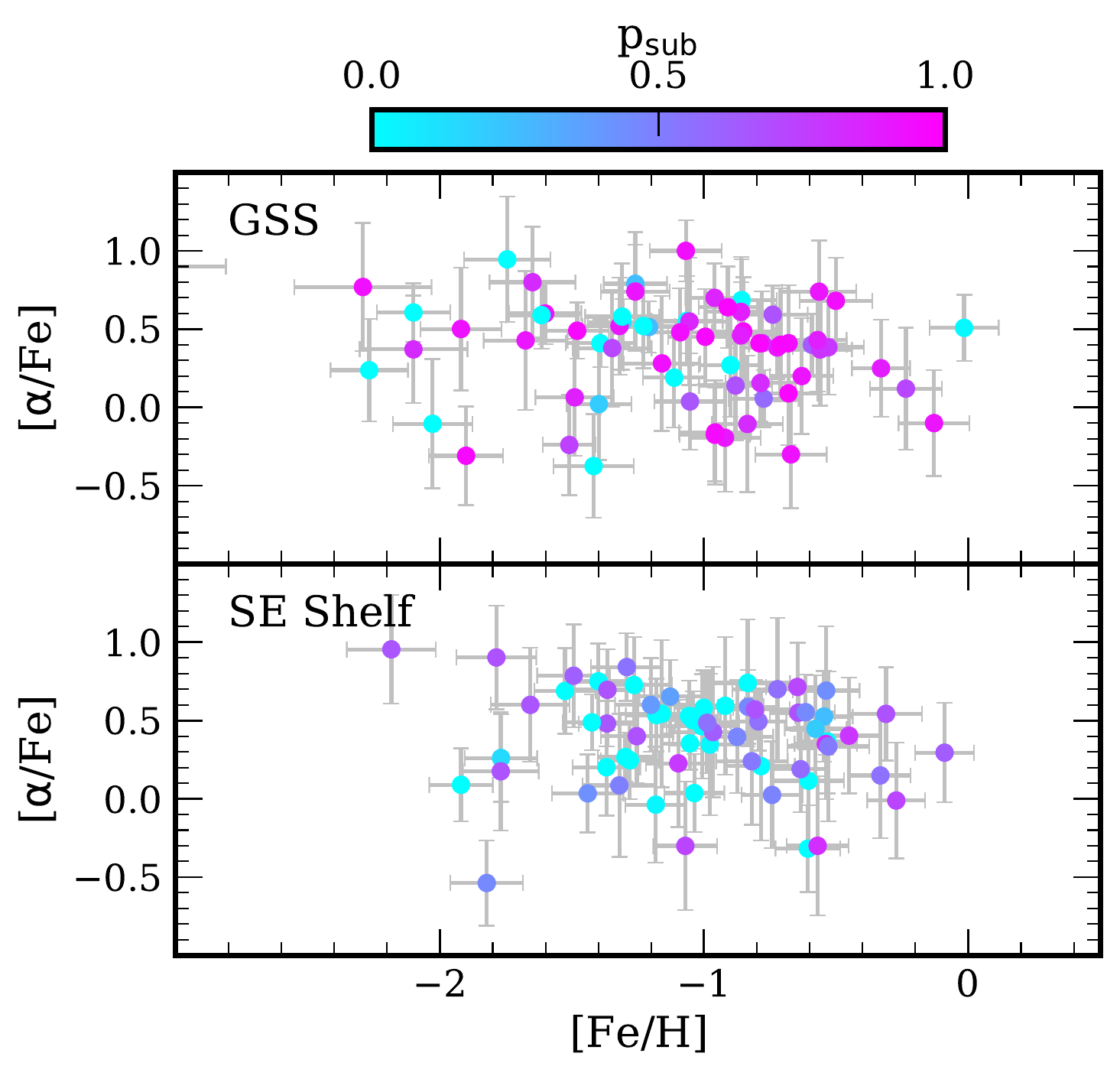}
    \caption{\feh\ versus \alphafe\ for M31 RGB stars in spectroscopic fields spanning the GSS (top panel) and SE shelf (bottom panel) (\citealt{Gilbert2019,Gilbert2020,Escala2020a,Escala2020b}; J.~Wojno et al. in preparation). Each star is color-coded by its kinematically-based probability of belonging to substructure (\S~\ref{sec:vel}), i.e., stars with $p_{\rm sub} > 0.5$ ($p_{\rm sub} < 0.5$) are likely associated with GSS-related tidal debris (the smooth halo).}
    \label{fig:alpha_vs_feh}
\end{figure}

Figure~\ref{fig:alpha_vs_feh} shows \alphafe\ versus \feh\ for RGB stars in spectroscopic fields targeting the GSS and SE shelf in M31's stellar halo (\citealt{Gilbert2019,Gilbert2020,Escala2020a,Escala2020b}; J.~Wojno et al., in preparation), where each star is color-coded by its probability of belonging to the given substructure component(s) present in each field. These final samples consist of M31 RGB stars with reliable stellar parameter and abundance measurements that do not show clear evidence of strong TiO absorption in their spectra. Such TiO stars are omitted from the final sample because we did not model absorption from the molecule when generating our grid of synthetic spectra. Furthermore, the size of a potential validation sample of TiO stars that could be used to evaluate the accuracy of these abundance measurements is currently limited. In order to select the final sample of unpublished measurements in the 58 kpc field, we employed our standard criteria ($\delta$\feh\ $<$ 0.4, $\delta$\alphafe\ $<$ 0.4, and well-constrained $\chi^2$ contours in each fitted parameter). The only exception is that we used a color cut ($(V-I)_0 < 2$) to exclude possible TiO stars from our final sample in this field, where we have shown that the majority of TiO stars have colors redder than this threshold (e.g., \citealt{Escala2020a}). Table~\ref{tab:fields} summarizes the chemical abundance properties of the GSS and SE shelf as probed at the locations of our spectroscopic fields.

\section{Chemical Abundance Gradients in the Giant Stellar Stream}
\label{sec:grad}

\begin{figure*}
    \centering
    \includegraphics[width=\textwidth]{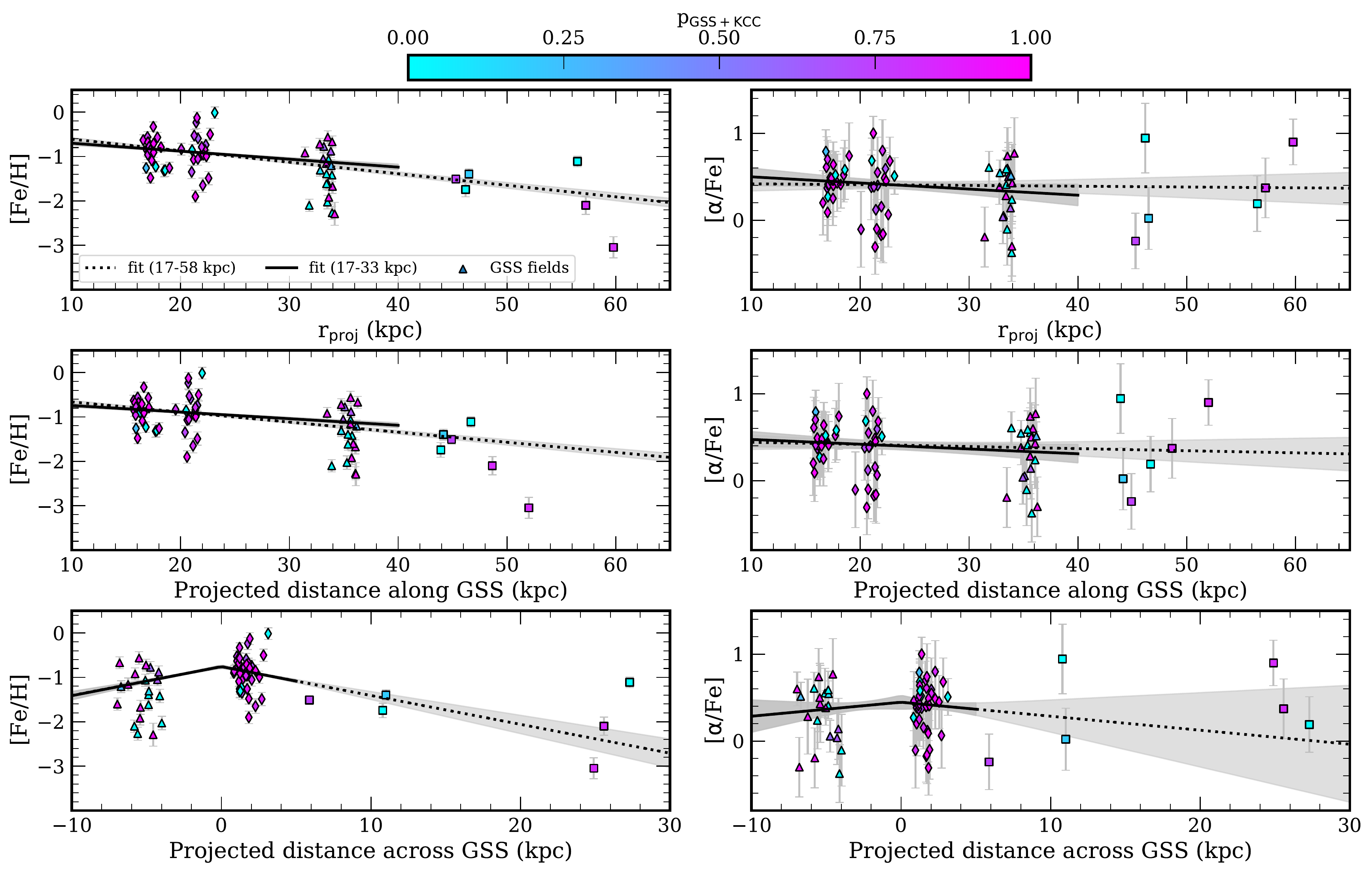}
    \caption{Spatial gradients of \feh\ (left column) and \alphafe\ (right column) in the GSS, where the GSS and KCC are treated as a single component. Data points correspond to abundance measurements for M31 RGB stars in spectroscopic fields spanning the GSS (Figure~\ref{fig:gss_2d}, Table~\ref{tab:fields}), where each point is color-coded according to its probability of belonging to any given substructure component present in a field. 
    Marker shape (triangle, diamond, square) denotes position across the GSS (eastern edge, core, and western envelope).
    Solid (dotted) lines and grey envelopes represent gradients measured considering only the inner halo GSS fields (17--33 kpc) and including the outer halo GSS fields (17--58 kpc).
    (Top row) Gradients measured as a function of projected distance from the center of M31. (Middle row) Gradients measured along an axis aligned with the high surface brightness core of the GSS, using the coordinate transformations defined by \citet{Fardal2006,Fardal2013}. (Bottom row) Gradients measured perpendicular to the GSS core. 
    The gradients are consistent between including and excluding the outer halo GSS stars.}
    \label{fig:gradients}
\end{figure*}

We measured spatial abundance gradients in the GSS from a sample of 62 M31 RGB stars with \feh\ and \alphafe\ measurements located in fields spanning the feature at 17, 22, and 33 projected kpc (Figure~\ref{fig:gss_2d}, Table~\ref{tab:fields}). As described in \S~\ref{sec:outer_halo_gradients}, we also considered the impact of a small sample of abundance measurements spanning the GSS in the outer halo at 45 and 58 projected kpc on the spatial gradients. We modeled the gradients by fitting a line to the data, allowing for uncertainties on both the dependent ($y$) and independent ($x$) axes (\S~\ref{sec:pos_err}). As opposed to describing the line by a slope ($k$) and intercept ($b$), we utilized the angle ($\phi = \tan^{-1}k$) and the orthogonal distance of the line from the origin ($b_\perp = b \cos \phi$) as model parameters. We used a Markov Chain Monte Carlo (MCMC) ensemble sampler \citep{Foreman-Mackey2013} to draw from the posterior probability distribution defined by the log likelihood under this model \citep{Hogg2010},

\begin{table*}
\centering
\begin{threeparttable}
    \caption{Spatial Abundance Gradients in the Giant Stellar Stream (17--33 kpc)}
    \begin{tabular*}{0.9\textwidth}{lcccccc}
    \hline \hline
    & \multicolumn{3}{c}{Slopes} & \multicolumn{3}{c}{Intercepts}\\
     \cmidrule(lr){2-4}\cmidrule(lr){5-7}
    & $r_{\rm proj}$ (kpc) & $m$ (kpc) & $|n|$ (kpc) & $r_{\rm proj}$ (kpc) & $m$ (kpc) & $|n|$ (kpc) \\
    \hline
    & \multicolumn{3}{c}{GSS+KCC} & \multicolumn{3}{c}{GSS+KCC}\\ \hline
    \feh    & $-$0.018 $\pm$ 0.003 & $-$0.015 $\pm$ 0.003 &  
    $-$0.065 $\pm$ 0.012 & $-$0.52 $\pm$ 0.08 & $-$0.59 $\pm$ 0.06 & 
    $-$0.76 $\pm$ 0.04\\
    \feh$_{\rm bias}$ & $-$0.026 $\pm$ 0.003 &  $-$0.021 $\pm$ 0.003 &  
    $-$0.081 $\pm$ 0.012 & $-$0.25 $\pm$ 0.08 & $-$0.36 $\pm$ 0.06 & 
    $-$0.61 $\pm$ 0.04\\
    \alphafe  & $-$0.007 $\pm$ 0.007 & $-$0.006 $\pm$ 0.006 &  
    $-$0.016 $\pm$ 0.024 & +0.57 $\pm$ 0.17 & +0.54 $\pm$ 0.14 & 
    +0.45 $\pm$ 0.08\\ \hline
    & \multicolumn{3}{c}{GSS} &  \multicolumn{3}{c}{GSS} \\ \hline
    \feh & $-$0.016 $\pm$ 0.004 & $-$0.013 $\pm$ 0.003 & 
    $-$0.053 $\pm$ 0.013 & $-$0.61 $\pm$ 0.09 & $-$0.68 $\pm$ 0.07 & 
    $-$0.83 $\pm$ 0.04\\
     \feh$_{\rm bias}$ & $-$0.022 $\pm$ 0.003 & $-$0.018 $\pm$ 0.003 & 
     $-$0.071 $\pm$ 0.012 & $-$0.36 $\pm$ 0.08 & $-$0.46 $\pm$ 0.07 & 
     $-$0.67 $\pm$ 0.04\\
    \alphafe & $-$0.006 $\pm$ 0.007 & $-$0.005 $\pm$ 0.006 &  
    $-$0.016 $\pm$ 0.026 & +0.52 $\pm$ 0.14 & +0.51 $\pm$ 0.16 & 
    +0.44 $\pm$ 0.10\\
    \hline
    \end{tabular*}
    \label{tab:gradients}
\begin{tablenotes}
\item Note. \textemdash\ Each row and column pair indicates the distance coordinate of the corresponding gradient slope and intercept for a given elemental abundance. 
The distance coordinates are given as projected radius ($r_{\rm proj}$), and distance along ($m$) and across ($n$) the GSS \citep{Fardal2006,Fardal2013}.
We measured gradients both including (top rows) and excluding (bottom rows) the KCC as part of the GSS. 
We also include values for \feh\ gradients measured by incorporating maximal bias estimates owing to selection effects (\S~\ref{sec:bias}). Note that the $|n|$-intercepts depend on the adopted zero-point of the GSS-aligned coordinate system.
\end{tablenotes}
\end{threeparttable}
\end{table*}

\begin{equation}
    \ln \mathcal{L} = -\frac{1}{2} \sum_{i=1}^{N} \left( \frac{\Delta_i^2}{\Sigma_i^2} - \ln (\Sigma_i) \right)
\end{equation}
\begin{equation}
    \Delta_i = y_i \cos\phi - x_i \sin\phi - b_\perp
\end{equation}
\begin{equation}
    \Sigma_i^2 = \frac{1}{p_{i,{\rm sub}}} \left( \delta y_i^2 \cos^2\phi + \delta x_i ^2  \sin^2\phi \right)
\end{equation}
where the index $i$ corresponds to a given RGB star with position $x_i$, abundance ratio $y_i$,  and associated uncertainties ($\delta x_i, \delta y_i$). We employed 10$^{2}$ walkers and 10$^{3}$ steps for a total of 5$\times$10$^{4}$ samples of each parameter when using the latter 50\% of each chain. We assumed flat priors on the model parameters ($\phi, b_\perp$) and incorporated the substructure probability ($p_{i, {\rm sub}}$) as an additional weighting term. Thus, the fitting procedure penalizes \feh\ and \alphafe\ measurements for RGB stars that are highly probable members of the kinematically hot stellar halo. Following the conclusion of the fitting procedure, we transformed the marginalized posterior probability distributions back to the more traditional ($k, b$) parameterization. We adopted the 50$^{\rm th}$ percentiles and 68\% confidence intervals of these distributions as the final values and uncertainties for each model parameter.

\begin{figure*}
    \centering
    \includegraphics[width=\textwidth]{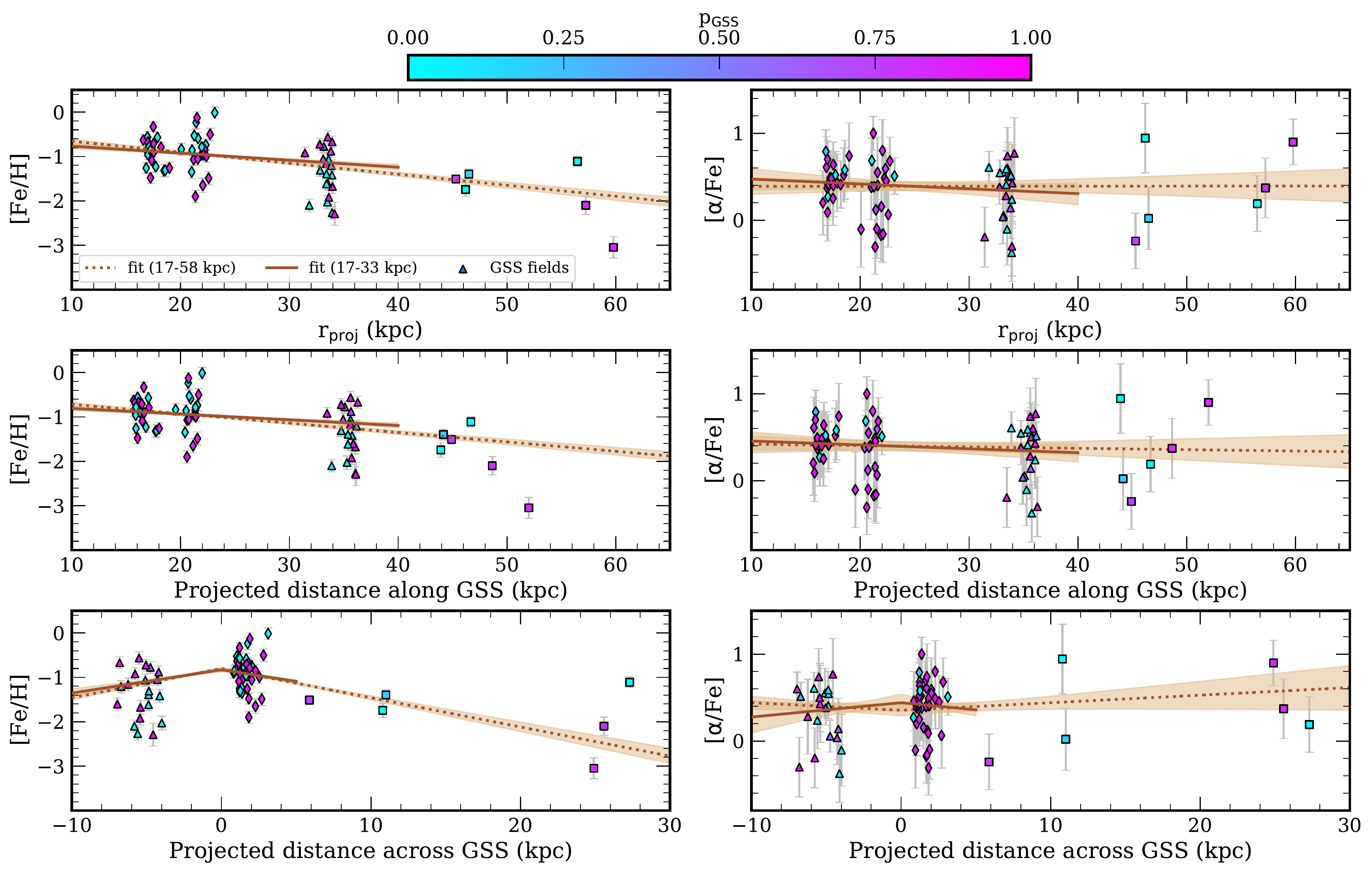}
    \caption{Same as Figure~\ref{fig:gradients}, except treating the GSS and KCC as separate components. All gradients measured between 17--33 projected kpc, which exclude the KCC, are statistically consistent with the case of including the KCC (Figure~\ref{fig:gradients}). The same is true for gradients measured between 17--58 projected kpc.}
    \label{fig:gradients_gssonly}
\end{figure*}

For our fiducial case, we fit for abundance gradients with respect to projected M31-centric radius ($r_{\rm proj}$) and defined $p_{i,{\rm sub}}$ as the probability that a RGB star belongs to \textit{any} substructure component, inclusive of the KCC. The physical motivation for this approach is the chemical similarity between the GSS and KCC, where current evidence suggests that their \feh\ and \alphafe\ distributions do not differ substantially between 17--22 kpc \citep{Gilbert2009,Gilbert2019,Escala2020a}. Our analysis provides further support for this conclusion, where we found that weighting gradient measurements solely toward RGB stars with a high probability of belonging to the GSS produces fully consistent results for the slopes. The gradient intercepts are marginally consistent (within $\sim$(1--2)$\sigma$) for \feh, where including the KCC results in more metal-rich values for the normalization, and are statistically consistent for \alphafe. The abundance gradient slopes, intercepts, and their uncertainties are presented in Table~\ref{tab:gradients}, where the top panels of Figures~\ref{fig:gradients} and~\ref{fig:gradients_gssonly} show the relationship between \feh\ and \alphafe\ and $r_{\rm proj}$ when including and excluding the KCC, respectively, as a contributor to the substructure probability. We measured a relatively steep, negative \feh\ gradient as a function of projected radius in the GSS, whereas we did not find evidence of a statistically significant (i.e., inconsistent with zero by at least 3$\sigma$) radial \alphafe\ gradient.

In order to distinguish between abundance gradients present along the high surface brightness core of the GSS and across the GSS envelope, we then transformed the M31-centric coordinates ($\xi, \eta$) for each RGB star into a GSS-aligned coordinate system ($m, n$) defined by \citet{Fardal2006, Fardal2013}. This system is described by the unit vectors $\hat{m} = (0.504, -0.864)$ (along the GSS core) and $\hat{n} = (-0.864, -0.504)$ (across the GSS envelope) in the ($\xi, \eta$) plane (Figure~\ref{fig:gss_2d}), where we have adopted ($\alpha_{\rm J2000}$, $\delta_{\rm J2000}$) = ($00^{\rm h}42^{\rm m}44^{\rm s}, +41^{\rm d}16^{\rm m}09.0^{\rm s}$) for the position of M31's center. \citet{Fardal2006} calculated these unit vectors from the slope traced out by the ($\xi, \eta$) sky positions of Canada-France-Hawaii Telescope (CFHT) imaging fields targeting the GSS core \citep{McConnachie2003}. We then shifted the center of the GSS-aligned coordinate system from $(m,n) = (0,0)$ to $(m,n) = (0, 0.34)$ degrees to correspond to the location of the transverse peak of GSS RGB star counts, which \citet{Fardal2013} determined from background subtracted imaging of M31's southeast quadrant \citep{Irwin2005}. We converted the $m$ and $n$ coordinates from degrees to kpc using a line-of-sight distance to M31 of 785 kpc \citep{McConnachie2005}. In the subsequent analysis, we present transverse gradients in terms $|n|$ (the absolute coordinate) as opposed to $n$ to clearly reflect trends between the GSS core and envelope, although we plot data points with respect to $n$ to preserve the spatial orientation of the GSS on the sky.

The middle (lower) panels of Figures~\ref{fig:gradients} and~\ref{fig:gradients_gssonly} show the resulting abundance gradients computed along (across) the GSS, and Table~\ref{tab:gradients} summarizes the relevant parameters.  As before, we did not detect statistically significant \alphafe\ gradients in either dimension of the GSS-aligned coordinate system. 
We detected negative \feh\ gradients both across and along the GSS. This former trend reflects a steep decline in the metallicity between the core and envelopes of the stream as previously observed in photometric metallicities \citep{Ibata2007,Gilbert2009}. If confirmed, the latter trend would represent the first detection of a significant spectroscopic \feh\ gradient along the stream, where this gradient is consistent with the radial \feh\ gradient within 1$\sigma$. Despite this similarity, it is unclear whether the apparent radial gradient is driven primarily by the gradient aligned with \textit{or} transverse to the GSS. In \S~\ref{sec:appendix}, we show that current data are consistent with the radial \feh\ gradient originating solely from an intrinsic \feh\ gradient in only one of these spatial dimensions, where larger samples are required to distinguish between these trends.

\subsection{The GSS in the Outer Halo}
\label{sec:outer_halo_gradients}


In order to explore chemical abundance trends in the GSS over a larger projected area, we expanded our analysis of gradients to include 6 M31 RGB stars with measurements of \feh\ and \alphafe\ (\citealt{Gilbert2020}; J.~Wojno et al. in preparation; Table~\ref{tab:fields}) present in spectroscopic fields beyond 40 projected kpc that are known to probe the GSS. Figures~\ref{fig:gradients} and~\ref{fig:gradients_gssonly} show the gradients measured between 17--58 projected kpc, which include the outer halo GSS stars, compared to our fiducial gradients measured between 17--33 projected kpc. Including the outer halo GSS stars results in \feh\ and \alphafe\ gradients with respect to $r_{\rm proj}$ and the GSS-aligned coordinate that are marginally consistent (within 1.6$\sigma$) with the parameters in Table~\ref{tab:gradients}. 
Although the sign of the transverse \alphafe\ gradient changes from negative to positive upon inclusion of the outer GSS stars, each case is consistent with a flat \alphafe\ gradient within the 1$\sigma$ uncertainties.
Thus, the incorporation of GSS stars beyond 40 kpc suggests that the declining trends of \feh\ with respect to projected distance across the GSS \textit{and} projected distance along the GSS continue out to the farthest positions probed by our data.


Additionally, we note that there is no statistically significant difference in the gradient slopes between including and excluding the KCC when measuring the gradients between 17--58 kpc, where this feature is not present in the line-of-sight velocity distributions (Figure~\ref{fig:vels}) of the 45 and 58 kpc fields. The gradient intercepts maintain marginal consistency (within (1-2)$\sigma$) regardless of inclusion of the KCC, where the most notable change occurs in the normalization of the transverse \feh\ gradient. In summary, larger samples spanning the GSS in the outer halo are necessary to confirm the identified trends from spectral synthesis based abundance measurements. We explore whether such trends with \feh\ persist in \fehphot\ measurements of probable GSS stars within our set of spectroscopic fields (Table~\ref{tab:fields}) in \S~\ref{sec:phot}.

\subsection{Photometric Metallicity Gradients in the GSS}
\label{sec:phot}

\begin{figure}
    \centering
    \includegraphics[width=\columnwidth]{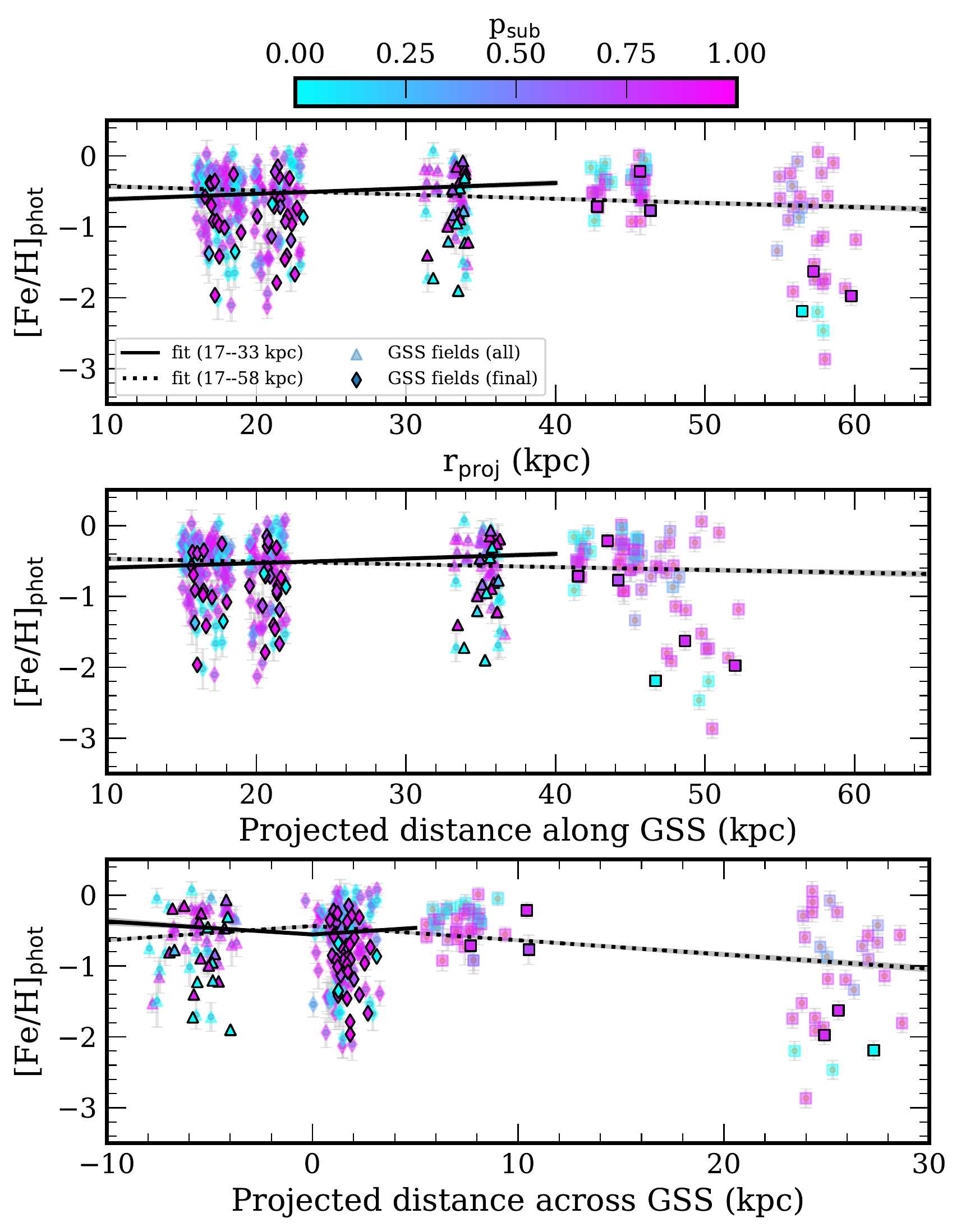}
    \caption{Photometric metallicity (\fehphot) gradients in GSS (\S~\ref{sec:phot}) measured from 
    all \ngssphot\ (\ngssouterphot) RGB stars spectroscopically identified as M31 members (\S~\ref{sec:vel}) across the GSS fields (Table~\ref{tab:fields}), assuming 9 Gyr PARSEC isochrones \citep{Marigo2017} with \alphafe\ = 0. We show the fitted gradients including (dotted line) and excluding (solid line) the outer halo GSS fields. RGB stars excluded from the final abundance sample (opaque points; \S~\ref{sec:abund_measure}) are shown as transparent points, where the marker shapes and color-coding are the same as Figure~\ref{fig:gradients}. The inferred metallicity trends differ between photometric and spectroscopic metallicity measurements even when controlling for selection effects, suggesting that the difference is intrinsic to the measurement methodologies.}
    \label{fig:fehphot_gradients}
\end{figure}

We further investigated spatial trends in the metallicity distribution of the GSS by repeating the above analysis using measurements of \fehphot\ for 
all \ngssphot\ (\ngssouterphot) spectroscopically identified 
RGB stars excluding (including) the outer halo GSS fields.
We measured \fehphot\ by interpolating the color and magnitude of each star on a grid of 9 Gyr PARSEC isochrones \citep{Marigo2017} with \alphafe\ = 0 as described by \citet{Escala2020a}. Figure~\ref{fig:fehphot_gradients} shows the spatial \fehphot\ gradients fit between 17--33 kpc and 17--58 kpc compared against \fehphot\ measurements for the final abundance sample (\S~\ref{sec:abund_measure}) and for all spectroscopically identified RGB member stars (\S~\ref{sec:vel}) across fields spanning the GSS (Table~\ref{tab:fields}). The abundance sample excludes TiO stars, which have high \fehphot, and includes only stars that have reliable \feh\ and \alphafe\ measurements (\S~\ref{sec:abund_measure}). The latter sample includes all RGB members regardless of whether they show spectral TiO signatures or have successful \feh\ and \alphafe\ measurements.

We chose to examine the \fehphot\ distribution of all RGB stars along the line-of-sight to the GSS, as opposed to solely RGB stars in the abundance sample, given that the former sample is larger and less subject to bias against stars with high \fehphot. The spatial \fehphot\ trends are qualitatively the same between all member stars and the abundance sample. The most notable difference is that the gradient intercepts are more metal-poor for the abundance sample, which omits TiO stars. We treated the GSS and KCC as the same substructure component, where disregarding the KCC as a contributor to the substructure probability does not result in a significant difference ($\gtrsim$1$\sigma$) in the inferred gradients with projected radius and along the GSS core when measured out to 33 or 58 kpc. 

As shown in Figure~\ref{fig:fehphot_gradients}, considering only the inner GSS sample yields marginally positive \fehphot\ gradients with respect to projected radius and projected distance along and across the GSS. 
However, when including RGB stars in the outer GSS, the 
\fehphot\ gradients become marginally negative. 
This suggests that the photometric metallicity along the GSS increases out to $\sim$30-40 projected kpc before decreasing at larger distances.  Furthermore, the photometry predicts a minor asymmetry in the metallicity distribution across the GSS core, where the eastern edge of the GSS appears to have higher \fehphot\ than both the core at $n \sim 1$ kpc and the extended western envelope.

The trends gathered from photometry are at odds with those derived from our spectral synthesis based metallicity measurements (e.g., Figure~\ref{fig:gradients}), which show consistently negative \feh\ gradients along and across the GSS.
Given that the \fehphot\ gradients measured from the abundance sample show the same qualitative behavior as the sample of RGB members and the photometric and spectroscopic \feh\ measurements are positively correlated, it is unlikely that biases incurred by selection effects (\S~\ref{sec:bias}) such as the omission of TiO stars can explain this discrepancy. The probable culprit is the disproportionately metal-rich disparity \added{(\fehphotspecdiffthirtythreekpcGSSfield\ $\pm$ \seomfehphotspecdiffthirtythreekpcGSSfield\ for the abundance sample)} between \fehphot\ and \feh\ measurements in the 33 kpc GSS field as compared to other fields. A possible explanation is that the necessary assumptions of constant stellar age and $\alpha$-enhancement to determine CMD-based metallicity estimates are particularly inappropriate for the stellar populations probed in this spectroscopic field. \added{In order to minimize this disparity, we would need to assume both older and $\alpha$-enhanced isochrones when measuring \fehphot\ for this field. Adopting $t = 14$ Gyr and \alphafe\ = +0.3,\footnote{The PARSEC isochrones do not have an $\alpha$-enhanced option. We chose this isochrone set despite this because of the need for stellar evolutionary models that include molecular TiO in M31's stellar halo. Thus, we estimated the effect of assuming $\alpha$-enhanced isochrones on \fehphot\ from \citet{Gilbert2014}. Using \citet{VandenBerg2006} models, they found that assuming \alphafe\ = +0.3 decreases \fehphot\ by $\sim$0.2 for M31 RGB stars. Given that this estimate is a relative quantity, it should not depend significantly on the adopted isochrone set.} as opposed to $t = 9$ Gyr and \alphafe\ = 0, reduces the difference between \fehphot\ and \feh\ by $\sim$0.3 to $\sim0.2-0.4$ dex. We emphasize that using different values for the stellar age and $\alpha$-enhancement will not necessarily resolve this metallicity discrepancy, given that the assumption of mono-age and mono-\alphafe\ stellar populations is unrealistic for GSS stars with a range of stellar ages \citep{Brown2006, Tanaka2010} and [$\alpha$/Fe] \citep{Gilbert2019,Escala2020a}.}

We refer the interested reader to \citet{Escala2020b} for a detailed discussion of the systematics between spectral synthesis and CMD-based metallicity measurements in the context of M31's stellar halo. We compare the spatial metallicity trends implied by both the spectroscopic and photometric metallicity measurements to those from the literature in \S~\ref{sec:lit}.

\subsection{Apparent Transverse vs. Aligned Metallicity Gradients}
\label{sec:appendix}

\begin{figure*}
    \centering
    \includegraphics[width=\textwidth]{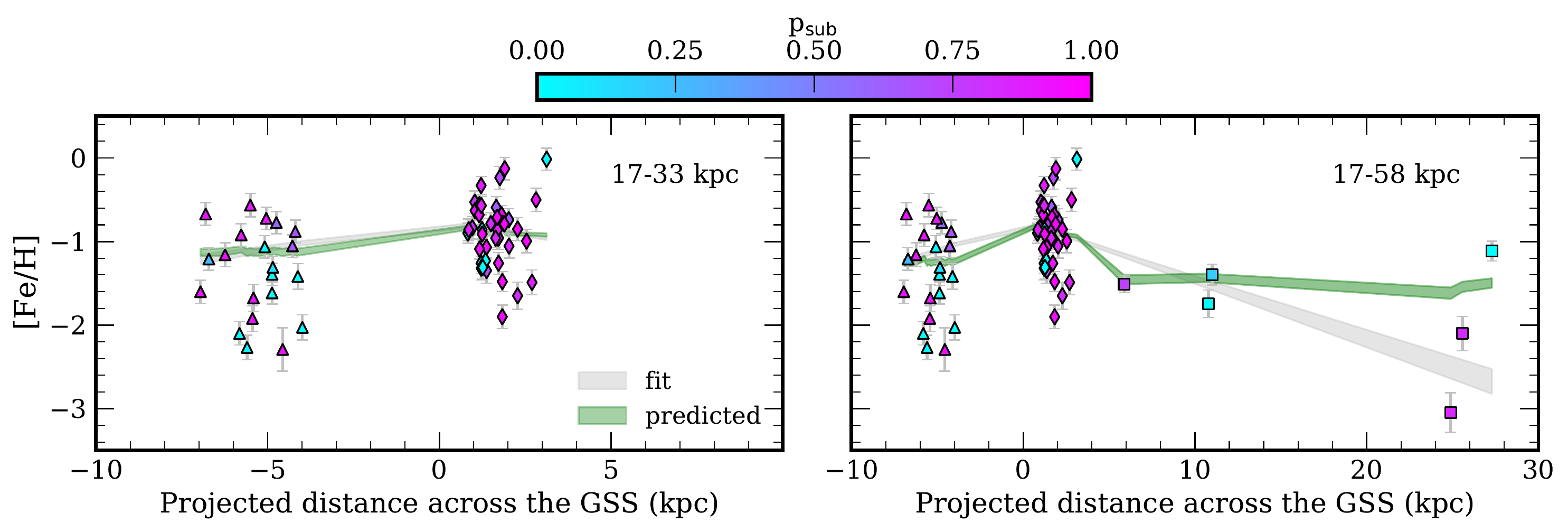}
    \caption{\feh\ gradient measured out to 33 kpc (left panel) and 58 kpc (right panel) as a function of projected distance across the GSS. The data points and color-coding are the same as Figure~\ref{fig:gradients}. The shaded gray envelopes are the 68\% confidence intervals of the fitted \feh\ gradients in the transverse direction (Table~\ref{tab:gradients}). The shaded green envelopes represent the apparent transverse gradient assuming that only an aligned gradient is present in the data (\S~\ref{sec:appendix}). Within both 33 kpc and 58 kpc, a star's projected distance along the GSS appears to be predictive of \feh\ regardless of its projected distance across the GSS, indicating that an intrinsic \feh\ gradient likely exists in only one of these coordinates, although larger samples are needed 
    to distinguish between these trends.}
    \label{fig:feh_n_grad}
\end{figure*}

In \S~\ref{sec:grad}, we measured \feh\ gradients out to 33 and 58 kpc  as a function of projected M31-centric radius ($r_{\rm proj}$), projected GSS-aligned distance ($m$), and projected absolute distance orthogonal to the GSS ($|n|$). The statistical consistency of the radial and $m$ gradients (Table~\ref{tab:gradients}) prompts the question of whether the observed radial gradients are primarily driven by the gradients along or across the GSS. 
The former (latter) case would indicate that there is little to no intrinsic transverse (aligned) \feh\ gradient, but rather that the observed transverse (aligned) gradient is an apparent consequence of an intrinsic GSS-aligned (GSS-transverse) gradient combined with the particular spatial sampling of the spectroscopic fields (Figure~\ref{fig:gss_2d}). Thus, we utilized 5$\times$10$^{4}$ pairs of transformed parameters sampled from the posterior probability distribution of our gradient model (\S~\ref{sec:grad}) to infer the expected behavior of an apparent transverse \feh\ gradient in the GSS by assuming that only an intrinsic aligned gradient is present. 

For each slope-intercept pair drawn from the aligned gradient model, we calculated the predicted \feh\ value at the observed GSS-aligned coordinate of each RGB star, then assigned this value to the corresponding transverse coordinate. Figure~\ref{fig:feh_n_grad} shows the 68\% confidence intervals for the  \feh\ values predicted from the aligned coordinates alone (green envelopes) compared to the ``true'' \feh\ values inferred from the transverse coordinates (gray envelopes). Within both 33 and 58 projected kpc, GSS-aligned position appears to predict \feh\ at a given transverse position, although it is less immediately clear in the 58 kpc case. This indicates that the observed radial gradient in our data is most likely driven by 
\textit{either} the gradient along \textit{or} across the GSS in the inner halo,
and tentatively the outer halo, with the caveat that larger samples in the outer halo are required to 
distinguish between these two trends.

\subsection{Relationship to the Southeast Shelf} \label{sec:se_shelf}


Motivated by the multiple lines of evidence for an association between the SE shelf and GSS \citep{Gilbert2007,Escala2020a}, we compared their chemical abundances, defining the sample for each feature from all stars in fields where it is present (Table~\ref{tab:fields}). We computed \fehavg\ and \alphafeavg\ for the GSS and SE shelf via 10$^{4}$ bootstrap resamplings of their abundance distributions, weighting by substructure probability and the inverse variance of the measurement uncertainty. Table~\ref{tab:gss_seshelf} presents the 50$^{\rm th}$ percentiles of the resulting distributions (and the associated uncertainties from the 16$^{\rm th}$ and 84$^{\rm th}$ percentiles), where we included halo stars in the 45 and 58 kpc fields in our calculations (\S~\ref{sec:outer_halo_gradients}). In summary, the mean chemical properties of the GSS and SE shelf agree within $1\sigma$, regardless of whether we include or exclude the KCC. There is tentative evidence that the SE shelf is both more metal-rich and $\alpha$-enhanced than the GSS, and furthermore that the KCC is more metal-rich than the GSS alone, but larger sample sizes are required to confirm these possibilities.

\begin{figure}
    \centering
    \includegraphics[width=\columnwidth]{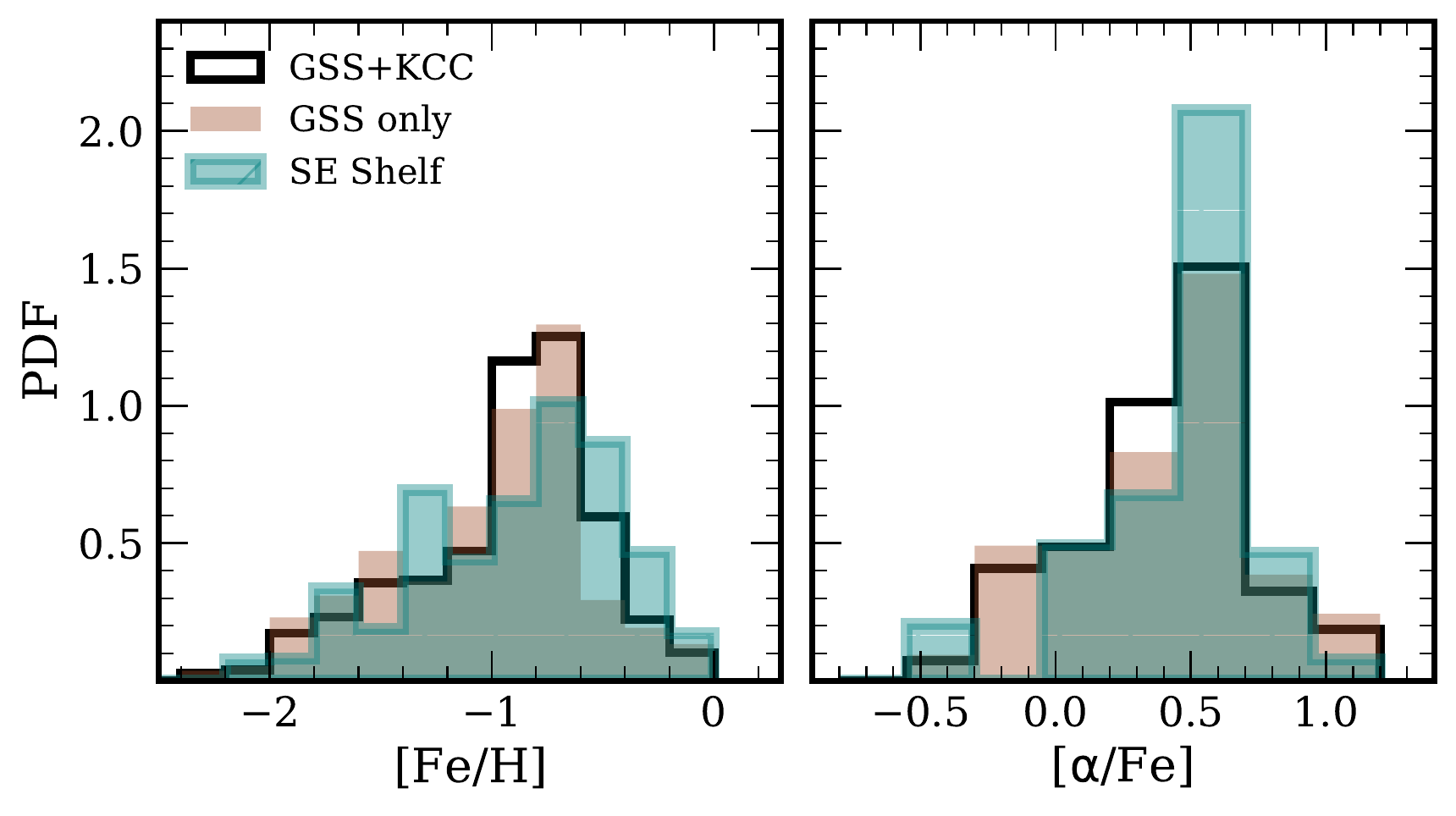}
    \caption{\feh\ (left) and \alphafe\ (right) distribution functions in the GSS, including (black outlined histogram) and excluding (shaded brown histogram) the KCC, and the SE Shelf (shaded blue histogram). 
    We adopted 0.20 (0.25) dex bins for \feh\ (\alphafe), which are comparable to the typical measurement uncertainty, and weighted the histograms by substructure probability and the inverse variance of the measurement uncertainty. The \feh\ and \alphafe\ distributions of the GSS and SE shelf appear to be consistent (\S~\ref{sec:se_shelf}).}
    \label{fig:se_shelf}
\end{figure}

Figure~\ref{fig:se_shelf} shows the \feh\ and \alphafe\ distribution functions for the GSS, both including and excluding the KCC, and the SE shelf. We constructed the histograms from all \feh\ and \alphafe\ measurements in spectroscopic fields known to contain a given feature (Table~\ref{tab:fields}), where we utilized substructure probability (\S~\ref{sec:vel}) and the inverse variance of the measurement uncertainty as weights. In order to evaluate whether the \feh\ and \alphafe\ distributions are statistically consistent between the GSS and SE shelf, we generated a distribution of $p$-values using the $k$-sample Anderson-Darling test. First, we selected all stars that are likely associated with a given feature ($p_{\rm sub}$ $>$ 0.5). We then perturbed the \feh\ and \alphafe\ measurements of these stars by 10$^{4}$ random draws from their Gaussian uncertainties and computed the test statistic between the GSS (including the KCC) and SE shelf for each iteration. We found that we could not reject the null hypothesis that \feh\ and \alphafe\ for the GSS and SE shelf are drawn from the same distribution at or below a 10\% significance level within the 1$\sigma$ confidence intervals. This is the case even when adopting a more stringent threshold for substructure membership ($p_{\rm sub} > 0.75$) or when excluding the KCC as a contributor to the GSS substructure probability.

Based on current measurements, it is therefore feasible for the SE shelf to originate from the same progenitor as the GSS. This is consistent with the finding by \citet{Gilbert2007} that the \fehphot\ distributions of M31 RGB stars kinematically associated with the GSS and SE shelf agree when correcting for contamination by the dynamically hot stellar halo, thereby bolstering support for the chemical similarity of the GSS and SE shelf. However, we acknowledge that this apparent similarity between the GSS and SE shelf may be complicated by the presence of spatial \feh\ gradients in the GSS, which originate in its progenitor (\S~\ref{sec:gss_prog}). Such large-scale \feh\ gradients in the GSS may therefore prohibit the existence of a clear \feh\ signature for debris related to the merger event, making it more difficult to definitively associate substructure such as the SE shelf with the GSS.


\subsection{Sources of Systematic Uncertainty}

\subsubsection{Sample Selection} \label{sec:bias}


The selection criteria for our final sample (\S~\ref{sec:abund_measure}) introduces two primary sources of potential bias into our abundance measurements owing to (1) the exclusion of red, presumably metal-rich stars with strong TiO absorption in their atmospheres ($b_{\rm TiO}$), and (2) signal-to-noise ratio limitations, which preferentially affect our ability to measure abundances for metal-poor stars ($b_{\rm S/N}$). As in \citet{Escala2020a,Escala2020b}, we assessed the impact of these sources of bias on our measured \feh\ gradients by shifting each \feh\ measurement in a given spectroscopic field by its corresponding maximal estimate for $b_{\rm TiO}$ + $b_{\rm S/N}$. We determined $b_{\rm TiO}$ from the $\langle$\fehphot$\rangle$ difference between all RGB stars and those in the final sample. We computed $b_{\rm S/N}$ from the $\langle$\feh$\rangle$ difference between RGB stars with successful \feh\ measurements (regardless of \alphafe) and the final sample.

We include the \feh\ gradients measured between 17--33 kpc with bias estimates (\feh$_{\rm bias}$) in Table~\ref{tab:gradients}. Regardless of whether RGB stars beyond 40 kpc are included, the \feh\ and \feh$_{\rm bias}$ gradient slopes are consistent within 2$\sigma$. For both excluding or including outer halo GSS stars, the \feh$_{\rm bias}$ gradient slopes agree within 1.7$\sigma$ regardless of whether we treat the GSS and KCC as a single component. 
However, the normalization of the \feh$_{\rm bias}$ gradients increases by $\sim$(2--2.5)$\sigma$ compared to the \feh\ gradients as a result of statistically taking into account the red, photometrically metal-rich TiO stars omitted from the final sample. Thus, we can conclude that our findings of negative metallicity gradient slopes with respect to projected radius and along and along the GSS
are relatively robust against the exclusion of TiO stars as the dominant source of bias.

The \alphafe\ gradients are unaffected by S/N limitations as a source of bias. However, they could be affected by the omission of relatively metal-rich TiO stars, assuming a correlation between \feh\ and \alphafe\ in the GSS, such that metal-rich stars tend to be less $\alpha$-enhanced. Based on current data, it is unclear if this trend is uniformly present among GSS stars in all spectroscopic fields (Figure~\ref{fig:alpha_vs_feh}). \citet{Escala2020a} did not find evidence of a statistically significant decline in \alphafe\ with respect to \feh\ for neither the GSS nor KCC in the 22 kpc field, whereas the characteristic ``knee'' feature in the \alphafe\ vs. \feh\ plane may be more apparent (at \feh\ $\sim-0.9$) in the 17 kpc field \citep{Gilbert2019}. \citet{Escala2020b} additionally found a visible decline in \alphafe\ with \feh\ in the 33 kpc field. It is therefore challenging to predict the net impact of excluding TiO stars on the \alphafe\ gradients for the GSS.

\subsubsection{Definition of GSS-Aligned Axes} \label{sec:pos_err}

\begin{table}
    \footnotesize
    \centering
    \caption{Chemical Properties of the GSS and SE Shelf (17--58 kpc)}
    \begin{tabular}{lcccc}
        \hline \hline
        Comp. & \fehavg & $\sigma$(\feh) & \alphafeavg & $\sigma$(\alphafe) \\ \hline
        GSS+KCC &  $-$0.96 $\pm$ 0.06 & 0.44 $\pm$ 0.05 &  0.40 $\pm$ 0.05 & 0.31 $\pm$ 0.03 \\
        GSS & $-$1.03 $\pm$ 0.07 & 0.47 $\pm$ 0.05 & 0.39 $\pm$ 0.06 & 0.34$\pm$ 0.04 \\
        SE Shelf & $-$0.89$^{+0.07}_{-0.08}$ & 0.47$^{+0.04}_{-0.05}$ & 0.45$^{+0.04}_{-0.05}$ & 0.29 $\pm$ 0.05 \\
        \hline
    \end{tabular}
    \label{tab:gss_seshelf}
\end{table}

To determine whether the abundance gradients are robust to different definitions of GSS-aligned coordinate axes, we re-measured the gradients 
while introducing positional uncertainty terms to the fitting procedure. We did this for all combinations of cases including and excluding the KCC, as well as including and excluding the outer halo GSS fields. By employing Gaussian fits to imaging data from \citet{McConnachie2003}, \citet{Font2006} found that 80\% (1.28$\sigma$) of the Stream's luminosity was contained within $\pm$0.25 degrees of the core. Thus, we propagated an error of $\delta \theta$ = 0.2$^\circ$ through our coordinate transformations ($m = \cos(\theta)\xi + \sin(\theta)\eta$, where $\theta \sim -149.8$ degrees east of north for our adopted coordinate system defined by \citeauthor{Fardal2006}),\footnote{The choice of the zero-point for the GSS-aligned axes does not affect the determination of the gradient slopes, although it alters the values of the intercepts.} which translates to median errors of $\delta m$ = 0.02 kpc and $\delta n$ = 0.07 kpc.
Incorporating these position-dependent errors results in gradient slopes and intercepts that are unchanged within the quoted uncertainties (Table~\ref{tab:gradients}).


\subsubsection{Distance Variations Along the GSS}

Early studies of resolved stellar populations in the GSS revealed the three dimensional structure of the stream, where the line-of-sight distance to the stream increases with increasing projected distance along the stream from the center of M31 \citep{McConnachie2003}. Given that we have assumed a constant distance modulus for all spectroscopic fields (\S~\ref{sec:abund_measure}), we assessed the impact of line-of-sight distance variations along the GSS on our measured abundance gradients. We adopted updated distances derived from the CMD position of the tip of the RGB along the GSS \citep{Conn2016}, as probed by the PAndAS survey. Similarly to \citeauthor{McConnachie2003}, \citeauthor{Conn2016} found that the line-of-sight distance to the GSS increases as a function of angular separation from M31, with a distance gradient of 20 kpc per degree over an angular extent of 6 degrees. 

Based on these values, the 22 kpc (33 kpc) GSS field (Figure~\ref{fig:gss_2d}; Table~\ref{tab:fields}) is located approximately \gssdistS\ (\gssdistathree) kpc behind M31, corresponding to an increase of \deltadmS\ (\deltadmathree) magnitudes compared to our assumed distance modulus.\footnote{We omitted the 17 kpc field from this analysis because it is separated from the 22 kpc field by $\sim$3 kpc in line-of-sight distance \citep{Conn2016}, corresponding to a difference in distance moduli of only $\sim$0.01 mag. We also excluded the GSS fields in the outer halo (Table~\ref{tab:fields}) because our fiducial gradients consider only the inner halo GSS fields.} This translates to a weighted average difference in the photometric effective temperature and surface gravity of $\Delta$\teffphot\ = \deltateffS\ $\pm$ \deltatefferrS\ (\deltateffathree\ $\pm$ \deltatefferrathree) K and $\Delta$\loggphot\ = \deltaloggS\ $\pm$ \deltaloggerrS\ (\deltaloggathree\ $\pm$ \deltaloggerrathree) dex, respectively, for M31 RGB stars present in the field, using 9 Gyr PARSEC isochrones \citep{Marigo2017} with \alphafe\ = 0. Assuming values of \teffphot\ and \logg\ corresponding to the increased heliocentric distance to the GSS yields $\Delta$\teff\ = \deltateffspecS\ $\pm$ \deltateffspecerrS\ (\deltateffspecathree\ $\pm$ \deltateffspecerrathree) K, $\Delta$\feh\ = \deltafehS\ $\pm$ \deltafeherrS\ (\deltafehathree\ $\pm$ \deltafeherrathree) dex, and $\Delta$\alphafe\ = \deltaalphaS\ $\pm$ \deltaalphaerrS\ (\deltaalphaathree\ $\pm$ \deltaalphaerrathree) dex (considering only statistical errors), where the abundance variations are well within our systematic uncertainties.

Thus, the $\sim$20 kpc difference in line-of-sight distance between our innermost and outermost GSS fields within 40 projected kpc has a negligible impact on our derived stellar parameters, and consequently, on our measured abundance gradients within this radial range. \citet{Gilbert2009} found comparable results regarding the impact of GSS distance variations on differences in the photometric metallicity between the core and envelope of the stream. Furthermore, \citet{Vargas2014b} performed a supporting analysis, in which they varied the assumed line-of-sight distance to M31 halo stars (by $\sim$150 kpc in either direction), and found that it does not alter spectral synthesis based abundance measurements within their uncertainties.  

\section{Discussion} \label{sec:discuss}

\subsection{Comparison to Previous Studies}
\label{sec:lit}

\begin{figure*}
    \centering
    \includegraphics[width=\textwidth, height=5in, keepaspectratio]{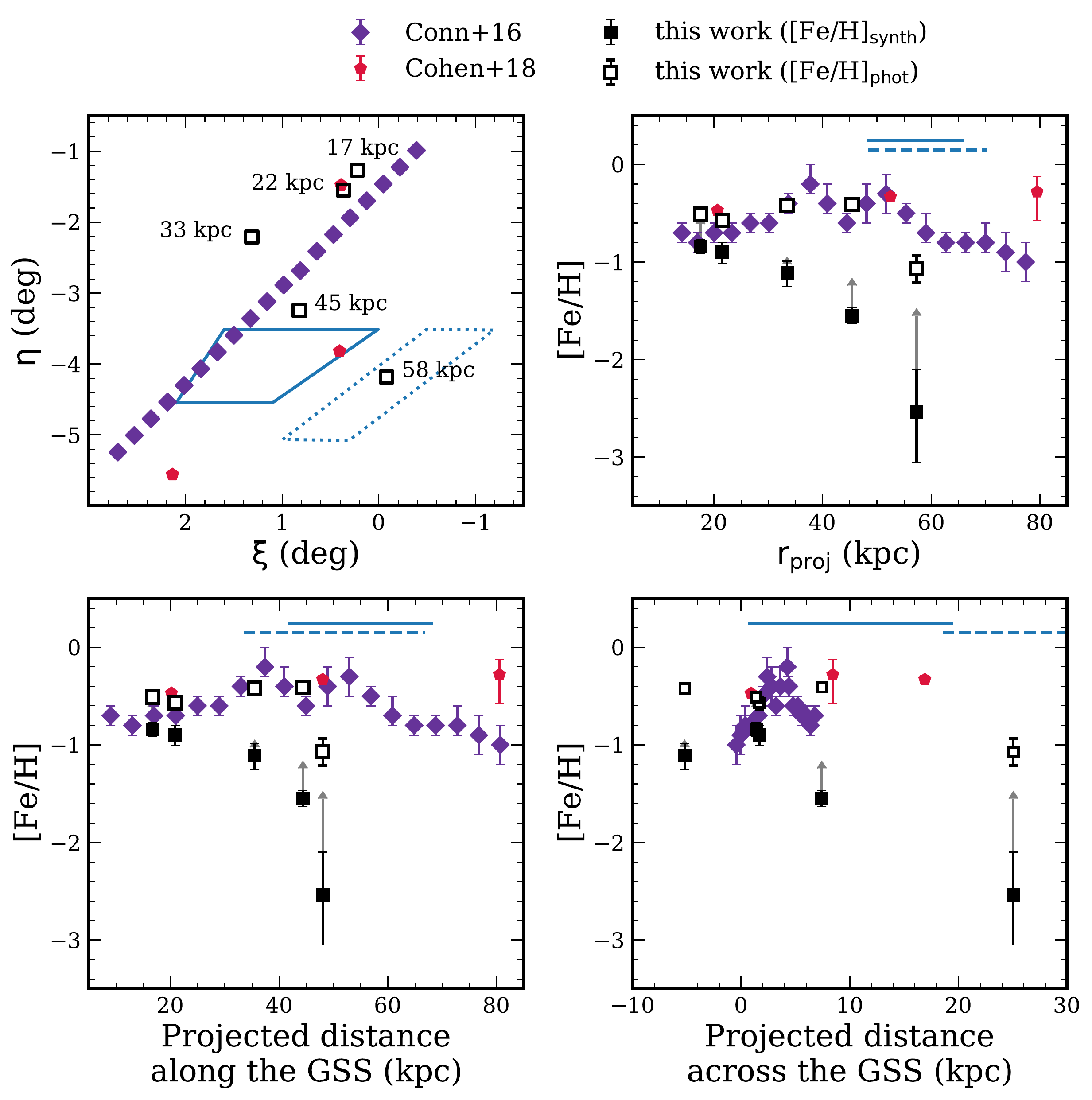}
    \caption{Comparison between this work and previous studies of spatial metallicity variations in the GSS (\S~\ref{sec:lit}). (Top left) Approximate locations of photometric (purple diamonds, red pentagons; \citealt{Conn2016,Cohen2018}) and spectroscopic (black open squares; this work, Table~\ref{tab:fields}) fields probing the GSS in M31-centric coordinates. The blue polygons denote the ``core'' (solid lines) and ``envelope'' (dashed lines) regions of the GSS from \citet{Ibata2007}. (Top right, bottom left, bottom right) \feh\ as a function of projected M31-centric radius, projected distance along the GSS, and projected distance across the GSS. The distance range spanned by the core (envelope) is shown in each panel as a solid (dashed) blue line. \feh\ refers to photometric metallicity for \citet{Conn2016} and \citet{Cohen2018}, whereas we show both spectral synthesis based (\feh$_{\rm synth}$; black filled squares; \S~\ref{sec:abund_measure}) and photometric (\fehphot; black open squares; \S~\ref{sec:phot}) metallicities for the spectroscopic fields. Gray arrows represent maximal bias estimates for \feh$_{\rm synth}$ in each field (\S~\ref{sec:bias}).}
    \label{fig:lit}
\end{figure*}

Early spectroscopic studies of individual RGB stars in the GSS at 22 and 33 projected kpc revealed a photometric metallicity difference of \kalirairajafehphotdiff\ dex \citep{Guhathakurta2006,Kalirai2006d}, supporting the possibility of metallicity variations in the stream as seen from photometry alone (\citealt{Ferguson2002}; \citealt{Ibata2007}; hereafter I07).
Using a large sample of photometric metallicities of spectroscopically confirmed GSS stars,  \citeauthor{Gilbert2009}\ (\citeyear{Gilbert2009}; hereafter G09) corroborated I07's core versus envelope metallicity dichotomy (top left panel of Figure~\ref{fig:lit}) by finding that GSS stars located at 17, 22, and 33 projected kpc near the core were more metal-rich by $\sim$\gilcoreandfehphotdiffmed\ (\gilcoreathirteenfehphotdiffavg\ $\pm$ \gilcoreathirteenfehphotdiffavgerr) dex than GSS stars located at 45 (58) projected kpc (without line-of-sight distance corrections). \citeauthor{Gilbert2009}\ concluded that their defined GSS core has an identical metallicity distribution to the 45 kpc field, and is significantly more metal-rich than the envelope as represented by the 58 kpc field.


The G09 fields are nearly identical to those utilized in this work, and target the same stellar populations. Indeed, with regard to \fehphot\ measurements, we find a difference of \fehphotcorevsand\ $\pm$ \fehphotcorevsanderr\ (\fehphotcorevsathirteen\ $\pm$ \fehphotcorevsathirteenerr) dex,\footnote{The details of sample selection are the most likely explanation for the slight discrepancy between this work and \citet{Gilbert2009} for the \fehphot\ difference between the G09 core fields and the 45 kpc field. We incorporated additional RGB stars published by \citet{Kirby2020}. Differences in the assumed isochrone age or model set should not significantly alter relative measures of \fehphot\ computed within a given data set. Although G09 considered only RGB stars within $\pm2\sigma_v$ of the GSS, in contrast to our usage of the KCC-inclusive substructure probability, upweighting likely GSS stars in our analysis would exacerbate the discrepancy because the KCC is more metal-rich than the GSS (Table~\ref{tab:gss_seshelf}).} such that the G09 core fields are nearly as metal-rich as the 45 kpc field (more metal-rich than the 58 kpc field). From our spectral synthesis based metallicity measurements, we find that the G09 core fields are more metal-rich than the 45 (58) kpc fields by \fehsynthcorevsand\ $\pm$  \fehsynthcorevsanderr\ (\fehsynthcorevsathirteen\ $\pm$ \fehsynthcorevsathirteenerr) dex, which at face value suggests a steeper decline in metallicity between the GSS core and envelope. The difference between the trends predicted by the photometric and spectroscopic metallicities cannot be accounted for by field-to-field variations in estimates of the \feh\ bias resulting primarily from the omission of red TiO stars (\S~\ref{sec:bias}; Figure~\ref{fig:lit}), which modifies the \feh\ difference between the core and 45 (58) kpc fields to \fehsynthcorevsandbias\ $\pm$ \fehsynthcorevsanderr\ (\fehsynthcorevsathirteenbias\ $\pm$ \fehsynthcorevsathirteenerr) dex.
However, when comparing results from various studies on spatial metallicity variations in the GSS, it is important to acknowledge varying definitions of the stream's core. For example, the G09 fields that define the GSS core are not spatially co-located with the core from I07 (top panels of Figure~\ref{fig:lit}), where the region spanned by the former (latter) covers $\sim$17--33 (48--66) projected kpc. 
Thus, the \fehphot\ difference examined by I07 primarily reflects orthogonal metallicity variations beyond 40 kpc in the GSS (bottom right panel of Figure~\ref{fig:lit}), whereas the spatial distribution of the G09 fields presents a more complex picture.

Figure~\ref{fig:lit} provides a view of metallicity variations in the GSS on equivalent spatial footing, as a function of projected radius, GSS-aligned distance, and GSS-transverse distance, while also placing our spectral synthesis based \feh\ measurements in the context of the literature \citep{Ibata2007,Conn2016,Cohen2018}. We substituted G09's \fehphot\ measurements with those from this work (\S~\ref{sec:phot}) for a similar set of spectroscopic fields (Table~\ref{tab:fields}) for the sake of homogeneity. We transformed the M31-centric coordinates of the imaging fields from \citet{Conn2016} (hereafter C16) and \citet{Cohen2018} (hereafter C18), and the area spanned by I07's core and envelope regions, into the GSS-aligned coordinate system of \citet{Fardal2006,Fardal2013} for direct comparison with our results.

First, we summarize the methodology and main results of the relevant photometric studies. C16 derived azimuthally averaged RGB metallicities spanning 70 projected kpc along the GSS by modeling PAndAS CMDs as a combination of weighted isochrones and a MW foreground contamination model \citep{Martin2013}.
C18 obtained CMD-based metallicities for individual RGB candidates in pencil-beam HST/ACS fields from Project AMIGA \citep{Lehner2020} and \citet{Brown2006} targeting the GSS at 21, 52, and 80 projected kpc. Neither C16 nor C18 correct for contamination of the GSS by M31's kinematically hot stellar halo, although they show that the influence of M31's halo on their results within 50 kpc should not be significant. Both studies found evidence for an increase in \fehphot\ with projected distance along the GSS out to $\sim$45--50 kpc, after which the behavior of \fehphot\ with GSS-aligned distance becomes less certain owing to heavy MW contamination.\footnote{C16 note that the MW contamination fraction in their outermost imaging subfields exceeds 80\% and may not therefore be representative. C18 similarly comment that the results for their 80 kpc field are highly sensitive to assumptions regarding their adopted MW foreground contamination model.} Thus, it is currently unclear whether CMD-based metallicites predict a plateau or a decline in the GSS-aligned gradient beyond $\sim$50 kpc. As for GSS-transverse distance, the range spanned by the C16 and C18 data is limited to that of the I07 core region, where the net \fehphot\ trend seems to be at most marginally positive.

Our \fehphot\ measurements broadly agree with C16 and C18 between $\sim$0--10 kpc across the GSS and within $\sim$45 kpc along the GSS (Figure~\ref{fig:lit}). However, our results diverge beyond this latter point, where we find an $\sim$\mfiftyfehphotdifflo--\mfiftyfehphotdiffhi\ dex lower average metallicity at $\sim$50 kpc. Potential reasons for this difference could be (1) unaccounted for contamination in the PAndAS/HST data by red MW dwarf stars with high inferred \fehphot, or (2) issues regarding sample selection and the associated Poisson noise in the sparse outer regions of the GSS. Although neither C16 nor C18 provide constraints in the I07 envelope region, the combination of these measurements with those from this work (and equivalently G09) appear to suggest that the ``edge'' of the photometrically metal-rich core occurs between $\sim$20--25 kpc across the GSS (see also C18). 
However, we have shown that it is unclear whether the core-envelope dichotomy visible from photometric metallicities clearly extends to spectroscopic metallicities based on currently available data (\S~\ref{sec:appendix}), where we cannot distinguish between an intrinsic gradient along or across the GSS.

Figure~\ref{fig:lit} also demonstrates that the \feh\ measurements show an apparent decline with projected distance along the GSS that is inconsistent with the qualitative trends predicted by \fehphot\ measurements in this work, C16, and C18. If the radial \feh\ gradient of the stream is intrinsic to the GSS-transverse distance (and not the GSS-aligned distance; \S~\ref{sec:appendix}), some of this inconsistency could result from our pencil-beam spectroscopic fields at 33, 45, and 58 projected kpc probing metallicity variations between the core and the envelope rather than those between the inner and outer GSS. However, this cannot entirely explain
the discrepancy between trends deduced from CMD-based and spectral synthesis based metallicities, given that it persists for the fields at 17 and 22 projected kpc near the GSS core. Thus, at least some of this discrepancy
is likely fundamental to the measurement metholodogies (\S~\ref{sec:phot}), where this interpretation is supported by the general similarity between \fehphot\ gradients from various studies. 
As we have previously discussed (\S~\ref{sec:outer_halo_gradients},~\ref{sec:appendix}), additional spectroscopy in the outer GSS is required to provide improved constraints on the stream's spatial abundance patterns.

\subsection{Implications for the GSS Progenitor}
\label{sec:gss_prog}

\begin{figure*}
    \centering
    \includegraphics[width=\textwidth]{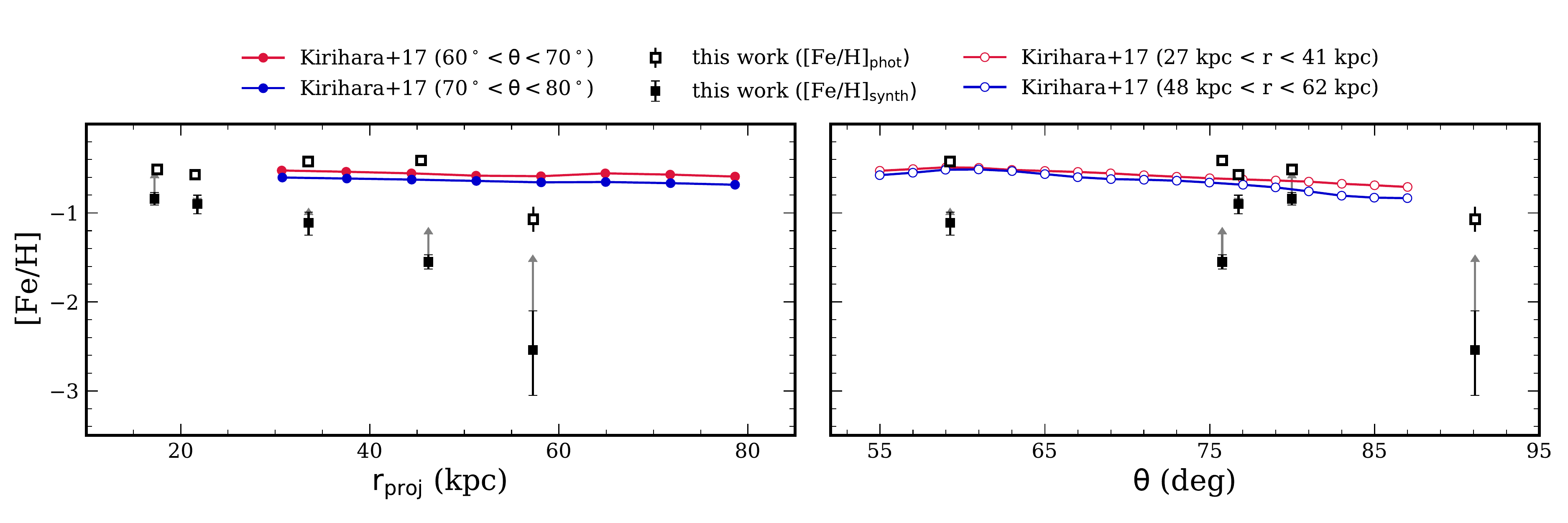}
    \caption{Comparison between observed (this work) and predicted \citep{Kirihara2017} radial (left) and azimuthal (right) metallicity variations in the GSS (\S~\ref{sec:gss_prog}). The \citeauthor{Kirihara2017} model assumes a minor merger with a GSS progenitor described by a thick disk of stellar mass $M_d = 7.3 \times 10^8 M_\odot$, central $\langle$\feh$\rangle$ = $-$0.5, and gradient of $\Delta$\feh\ = $-$0.5 in units of disk scale length. The azimuthal angle $\theta$ is defined such that $\theta = 0^\circ$ is east and the GSS core is located at $\theta \sim 65^\circ$. The left panel shows predicted \feh\ trends for both the core (red points) and envelope (blue points) of the GSS, whereas the right panel shows \feh\ trends for inner (open red points) and middle (open blue points) radial regions of the GSS. We include both spectral synthesis based (\feh$_{\rm synth}$) and CMD-based (\fehphot) metallicities for our spectroscopic fields. Gray arrows represent maximal bias estimates for \feh$_{\rm synth}$ in each field (\S~\ref{sec:bias}).}
    \label{fig:kiri17}
\end{figure*}

Both major and minor merger models for the formation of the GSS broadly reproduce the observed morphological and kinematical features of the stream and its associated shells (\citealt{Fardal2006,Fardal2007,Fardal2008,Fardal2013,MoriRich2008,Sadoun2014,Kirihara2014,Kirihara2017,Miki2016} for minor mergers; \citealt{Hammer2010,Hammer2018,DSouzaBell2018} for major mergers). Among minor merger models, rotating, disky progenitors better match the observed asymmetric structure of the GSS than spheroidal counterparts \citep{Fardal2008,Fardal2013,Kirihara2017}, although neither class of progenitor models can currently account for the existence of the KCC \citep{Gilbert2019} or the disturbed nature of M31's disk (e.g., \citealt{Dorman2015,Bernard2015,Williams2015}). 
To first order, major merger models explored thus far can simultaneously explain M31's disk and halo properties, though this does not necessarily disqualify a minor merger from being responsible for the GSS's formation. 

Thus, it is currently unknown whether the GSS progenitor had a stellar mass of $(1 - 5) \times 10^9 M_\odot$, or $\sim10^{10} M_\odot$, as respectively predicted by minor and major merger models (see above references). Current observational constraints on the stellar mass of the GSS progenitor from chemical abundance measurements place it between that of the LMC and M32 ($(1 - 5) \times 10^9 M_\odot$; \citealt{Gilbert2019}) when correcting for potential sources of observational bias (\S~\ref{sec:bias}),
which is consistent with predictions of minor merger models for the formation of the GSS. However, \citeauthor{Gilbert2019} caution that this cannot be interpreted as direct evidence in favor of a minor merger scenario without knowledge of where the GSS stars originate from in the progenitor, if the progenitor possessed a metallicity gradient. Along with prior studies (\S~\ref{sec:lit}), this work has shown that this situation is indeed the case given the observed presence of spatial metallicity gradients in the GSS.

Simulations of minor and major merger scenarios for the formation of the GSS that track stellar metallicity ubiquitously predict the existence of strong gradients in the progenitor in order to approximately match observations \citep{Fardal2008,MoriRich2008,Miki2016,Kirihara2017,Hammer2018,DSouzaBell2018}. Nonetheless, they differ in the details regarding the exact magnitude of the gradient (but less so in its direction; c.f. \citealt{Miki2016}) and the original location in the progenitor of GSS core stars. For example, some simulations posit that the GSS core is constituted by stars originating near the metal-rich center of the progenitor \citep{Fardal2008,Miki2016,Kirihara2017}, whereas others postulate that the stream debris comes from more metal-poor regions corresponding to a larger radial range within or the outskirts of the progenitor \citep{MoriRich2008,Hammer2018,DSouzaBell2018}. Thus, an understanding of how the distribution of GSS-related tidal debris on the sky maps to galactocentric radius in the progenitor is crucial for reconstructing the progenitor's metallicity gradient---and subsequently its average metallicity and inferred stellar mass---from available observational data.

Although the lack of a consensus on the original location of GSS stars in the progenitor limits our ability to directly constrain its metallicity gradient, comparisons between current model predictions and data are informative for identifying potential areas of disagreement. Figure~\ref{fig:kiri17} shows CMD-based (\fehphot) and spectral synthesis based (\feh) metallicity measurements for the GSS in our spectroscopic fields (Table~\ref{tab:fields}) as a function of projected radius and azimuthal angle (defined such that $0^\circ$ is east and the GSS core is located at $\sim65^\circ$) alongside trends from the models of \citet{Kirihara2017}. Their model assumes a minor merger with a GSS progenitor described by a rotating thick disk with a stellar mass of $7.3 \times 10^8 M_\odot$, a central value of $\langle$\feh$\rangle$ = $-$0.5, and a gradient of $\Delta$\feh\ = $-$0.5 in units of disk scale length (where $R_d = 1.1$ kpc). \citeauthor{Kirihara2017}\ investigated the metallicity patterns in their simulated GSS analog, which resulted from the initial gradient in the progenitor, predicting that the strongest metallicity variations were azimuthal and located at large projected radii (48--62 kpc). Furthermore, stronger gradients in their model translated to more pronounced metallicity differences between the GSS core and envelope, and metallicity differences along the stream were most prominent in its innermost regions. 

Although the above scenario could be qualitatively consistent with our measurements, Figure~\ref{fig:kiri17} clearly illustrates that this model is not able to provide a quantitative match. Considering only our fields within 33 kpc, the predicted trends are generally too metal-rich for the \feh\ measurements, even when taking into account \feh\ bias terms (\S~\ref{sec:bias}), although it is more similar to the equivalent \fehphot\ measurements. Additionally, the observed azimuthal behavior of \feh\ is more complicated than can be accounted for by the model. Although the former discrepancy could be minimized by assuming a more metal-poor center for the progenitor, a similar effect could presumably be achieved if the GSS core originates from further out in the progenitor's disk than is the case in this model. Additionally considering fields out to 58 kpc highlights the fact that the observed radial metallicity gradient may be much steeper than that predicted by this model, which could indicate a need for a stronger initial gradient in the progenitor.

Given that few GSS formation models that track stellar metallicity take the additional step of quantifying the predicted abundance ratios \citep{Fardal2008,Miki2016,Kirihara2017}, it is unclear if they can generally reproduce sufficiently strong gradients in comparison to our spectroscopic and photometric metallicity measurements. From a statistical sample of major merger scenarios for M31's formation, \citet{DSouzaBell2018} found that tidal debris from GSS progenitor analogs exhibited metallicity variations as large as 1 dex, but did not further quantify such results.\footnote{A limitation of current (cosmological, hydrodynamical) simulations that explore major merger scenarios for the GSS's formation \citep{DSouzaBell2018,Hammer2010,Hammer2018} is that they are computationally expensive, such that the resolution is necessarily lower than that of N-body minor merger models. Thus, kinematical and chemical structure cannot be resolved in these simulations at the level of detail dictated by the observations.} Furthermore, although this class of simulations demonstrate core-envelope dichotomies \citep{Fardal2008,MoriRich2008,Kirihara2017,DSouzaBell2018}, they do not generally predict observed gradients {\it along} the stream, as may exist in our data. This is excepting the models of \citet{Miki2016}, which produced negative radial gradients of approximately $-0.01$ dex kpc$^{-1}$ (compared to $-0.018 \pm 0.003$ dex kpc$^{-1}$; Table~\ref{tab:gradients}). In the case of \citet{Fardal2008}, the initial gradient in the progenitor is calibrated to the results of \citet{Ibata2007}, as opposed to being set by the relationship between its stellar mass and metallicity. In general, current GSS formation models appear to be capable of generating the morphological structure of the stream and its associated shells despite assuming a wide range of mass and metallicity properties for the progenitor (e.g., \citealt{Hammer2018}), therefore limiting the predictive power of any given modeled metallicity gradient for the GSS.

Additional studies that perform detailed modeling of the GSS metallicity distribution and careful comparisons to observations are therefore needed. In particular, models that also track $\alpha$-elements will be instructive. The lack of significant spatial \alphafe\ gradients in the GSS (\S~\ref{sec:grad}) suggests that its progenitor may have been uniformly $\alpha$-enhanced, or that its \alphafe\ variations are below the detectable threshold set by our typical measurement uncertainty (i.e., $\lesssim$ 0.3). \added{The presence of spatial \alphafe\ variations in Local Group dwarf galaxies, such as MW dwarf spheroidal satellite galaxies and the Magellanic Clouds (e.g., \citealt{Kirby2011,Nidever2020}), has generally not been quantified, thus largely precluding comparisons of observational expectations for [$\alpha$/Fe] gradients to the GSS. The exceptions include chemical abundance studies of M31 satellite dwarf galaxies \citep{Vargas2014a}, where no strong evidence for significant \alphafe\ gradients was found, and Sgr \citep{Hayes2020}. \citeauthor{Hayes2020} 
measured an \alphafe\ gradient of \sgrcorestreamalphagrad\ (or \sgrcorestreamalphadiff\ in absolute difference) between the Sgr core and Sgr streams--in addition to weaker internal \alphafe\ gradients in the streams--that they interpreted as reflecting Sgr's \feh\ gradient combined with the characteristic anticorrelation between \alphafe\ and \feh\ in dwarf galaxies (e.g., \citealt{Shetrone2001,Venn2004,Kirby2011}). Given that the GSS possesses a significant \feh\ gradient (Table~\ref{tab:gradients}) and shows evidence for a decline in \alphafe\ with \feh\ in some spectroscopic fields (\S~\ref{sec:bias}), the GSS could therefore feasibly exhibit [$\alpha$/Fe] variations of $\lesssim$ 0.3 dex.}

Regardless of whether an \alphafe\ gradient exists in the GSS, its high average $\alpha$-enhancement (+0.40 $\pm$ 0.05; Table~\ref{tab:gss_seshelf}) can provide constraints on the nature of the GSS progenitor, and thus formation scenarios for the stream. From the first \alphafe\ measurements of individual RGB stars in the GSS, \citet{Gilbert2019} concluded that the GSS progenitor must have formed stars efficiently enough to enrich to high metallicity (\feh\ $\sim$ $-$0.9) before experiencing a precipitous decline in its star formation rate such that the yields of Type Ia supernovae dominated over those of core-collapse supernovae. Indeed, the GSS progenitor must have had more efficient star formation than that of the present-day massive dwarf galaxies of the Local Group (\citealt{Hasselquist2017,Mucciarelli2017} for Sagittarius; \citealt{Pompeia2008,Lapenna2012,VanderSwaelmen2013,Nidever2020} for the Magellanic Clouds) or even the dominant progenitor of the Milky Way's stellar halo (Gaia-Enceladus-Sausage; \citealt{Helmi2018,Haywood2018,Naidu2020}), which have $\langle$\alphafe$\rangle$ $\lesssim$ +0.2 dex. The observed average $\alpha$-enhancement of the GSS--and by extension, its progenitor--is therefore unusual compared to expectations of its stellar mass from dynamical modeling in a minor merger scenario ($M_\star$ $\sim$ (1$-$5) $\times$ 10$^{9}$ $M_\odot$).

Even a scenario in which a massive progenitor dwarf galaxy ($M_\star \sim 10^9 M_\odot$) is accreted sufficiently early to truncate its star formation history on the high \alphafe\ plateau (e.g., \citealt{Johnston2008,Lee2015}) is unlikely to explain the observed $\langle$\alphafe$\rangle$ of the GSS\@. Minor merger models for the GSS place its first pericentric passage and accompanying formation of the stream at $\lesssim$1 Gyr ago \citep{Fardal2006,Fardal2007,Fardal2008,Fardal2013,MoriRich2008,Kirihara2014,Miki2016}, where the cosmologically motivated models of \citet{Sadoun2014} time the initial accretion of the progenitor at $\sim$3 Gyr ago. In addition, the most recent star formation in the GSS occured $\sim$4 Gyr ago, where the GSS has a typical stellar age of $\sim$8 Gyr \citep{Brown2006}. Thus, a lower mass progenitor would have produced lower $\langle$\alphafe$\rangle$ than is observed in the GSS: the progenitor's star formation would have quenched
via interaction with M31's ionized circumgalactic medium \citep{Lehner2020} only within the last few Gyr, providing sufficient time for Type Ia supernovae to deplete \alphafe\ with respect to \feh. The extended star formation history of the GSS similarly constrains the scenario of a high mass progenitor ($M_\star \sim 10^{10} M_\odot$), although the interaction between M31 and the progenitor can begin as early as $\gtrsim$ 5--10 Gyr ago in this case \citep{DSouzaBell2018,Hammer2018}.

The most significant difference between a major versus minor merger scenario for the average $\alpha$-enhancement of the GSS is therefore not the accretion time of the event, but rather the ability of the progenitor to sustain efficient star formation--and high $\langle$\alphafe$\rangle$--over many Gyr such that the progenitor could simultaneously enrich to high \feh\ (up to at least $-$0.96 dex; Table~\ref{tab:gss_seshelf}). Simulations have shown that massive, star-forming Milky Way like galaxies (M$_\star \sim 10^{(9.7-10.7)} M_\odot$) can produce \alphafe\ $\sim$ 
+0.4 dex at \feh\ $\sim$ $-1$ dex \citep{Naiman2018,Mackereth2018,Gebek2021}, in broad agreement with the abundance ratios observed in the GSS.
On the observational front, \citet{Galazzi2021} recently presented the first measurements of \alphafe\ in star-forming, massive galaxies ($M_\star \sim 10^{(9.5-11.5)} M_\odot$) beyond the Local Group using $>$110,000 $z=0$ galaxies in SDSS DR7 \citep{sdssdr72009}, confirming that the positive correlation between \alphafe\ and stellar mass observed for quiescent massive galaxies (e.g., \citealt{Thomas2005,Galazzi2006,Conroy2014,Segers2016}) extends to these systems. However, \citeauthor{Galazzi2021}\ found that star-forming massive galaxies tend to have lower SFH-integrated \alphafe\ at a given stellar mass, with a mean value of \alphafe\ $\sim$ +0.15 dex at $M_\star \sim 10^{10.5} M_\odot$ and 1$\sigma$ upper limits of $\sim$ +0.3 dex. Assuming that the average $\alpha$-enhancement of the GSS is representative of the progenitor,\footnote{Major merger models for the GSS's formation predict that the GSS has significant contributions from the more metal-poor outskirts of the progenitor \citep{Hammer2018,DSouzaBell2018}. Assuming that \alphafe\ declines at high \feh, it is therefore possible that the GSS stars are biased toward higher \alphafe\ relative to the progenitor as a whole.} the GSS progenitor would be within $\lesssim$1.5$\sigma$ of this relation in a major merger scenario (with the caveat that the progenitor halted star formation at $z \sim 0.4$, although it was star-forming at the time of accretion). We therefore conclude that a massive GSS progenitor ($M_\star \sim 10^{10} M_\odot$) provides a more natural framework for explaining the high $\alpha$-enhancement and metalliity gradient of the GSS.


\section{Summary} \label{sec:summary}

The Giant Stellar Stream (GSS; \citealt{Ibata2001}) is the most prominent tidal structure in M31, covering a significant portion of its southeastern quadrant and likely polluting much of its stellar halo (e.g., \citealt{Brown2006,Richardson2008,Gilbert2009}).
Until recently, studies of the GSS's chemical composition were limited to photometric and calcium triplet based metallicity estimates, where \citet{Gilbert2019} presented the first \feh\ and \alphafe\ abundances in the stream. From an existing sample of \ntotgss\ RGB stars with measurements of \feh\ and \alphafe\ from the Elemental Abundances in M31 survey \citep{Escala2019,Escala2020a,Escala2020b,Gilbert2019,Gilbert2020,Kirby2020,Wojno2020},  we have investigated the two-dimensional chemical abundance distribution of the GSS from a set of spectroscopic fields (Table~\ref{tab:fields}) spanning 17--33 projected kpc (\S~\ref{sec:grad}). We have expanded this data set to include \feh\ and \alphafe\ measurements for 6 additional RGB stars in the western envelope of the GSS (\citealt{Gilbert2020}; J.~Wojno et al., in preparation) in order to extend our analysis beyond 40 kpc (\S~\ref{sec:outer_halo_gradients}). We have measured a pronounced negative \feh\ gradient ($-0.018 \pm 0.003$ dex kpc$^{-1}$; Table~\ref{tab:gradients}) and a negligible \alphafe\ gradient as a function of projected radius in the GSS. Although limited by sample size, the outer GSS data supports a continuation of the inner GSS abundance trends. We have also shown that the measured \feh\ and \alphafe\ gradients are largely insensitive to whether the GSS and the KCC \citep{Kalirai2006d,Gilbert2009,Gilbert2019} are treated as a single feature, suggesting that they indeed share a common origin.

The spectroscopic metallicity measurements show evidence for an apparent negative gradient between the inner and outer GSS along an axis defined by the high surface brightness core of the GSS, although it is unclear if this trend is a manifestation of intrinsic metallicity variations between the core and the envelope of the GSS combined with the spatial sampling of the spectroscopic fields (\S~\ref{sec:appendix}).
Recent photometric metallicity measurements of the GSS show evidence for a \textit{positive} gradient over a similar radial range \citep{Conn2016}. By measuring the photometric metallicity for  \ngssouterphot\ RGB stars in our spectroscopic fields spanning the GSS (\S~\ref{sec:phot}), we have confirmed that \fehphot\ trends in our data are similar to the literature (\S~\ref{sec:lit}) and thus conclude that differences between metallicity patterns predicted by spectroscopic and photometric measurements are likely intrinsic to the measurement methodologies. 

Although we do not detect a significant \alphafe\ gradient in the GSS, the high average $\alpha$-enhancement of the feature ($\langle$\alphafe$\rangle$ = +0.40 $\pm$ 0.05; Table~\ref{tab:gss_seshelf}) argues in favor of an origin in a major merger (M$_\star \sim 10^{10} M_\odot$), as opposed to a minor merger (M$_\star \sim 10^{9} M_\odot$), when combined with constraints regarding its star formation history \citep{Brown2006} and relatively high mean metallicity ($\langle$\feh$\rangle$ = $-$0.96 $\pm$ 0.06; Table~\ref{tab:gss_seshelf}). A massive, disky, star-forming galaxy could enrich to high \feh\ and \alphafe\ (e.g., \citealt{Galazzi2021}) by maintaining a high efficiency of star formation for many Gyr (\S~\ref{sec:gss_prog}).

In addition, we have demonstrated that the \feh\ and \alphafe\ distributions of the GSS are statistically consistent with those of the Southeast shelf (\S~\ref{sec:se_shelf}; Table~\ref{tab:gss_seshelf}), a tidal feature predicted by GSS formation models \citep{Fardal2006,Fardal2007} and subsequently discovered from spectroscopy \citep{Gilbert2007}, thereby providing support for a common origin scenario. However, metallicity gradients originating in the progenitor are a common feature of GSS formation models \citep{Fardal2008,MoriRich2008,Miki2016,Kirihara2017,Hammer2018,DSouzaBell2018}, such that it is unclear how an initial gradient translates to an observed gradient among the tidal debris (\S~\ref{sec:gss_prog}), thus limiting the ability to make chemical connections between features. Future advances in understanding the abundance patterns of the GSS will be instigated by larger samples of \feh\ and \alphafe\ measurements in the outer GSS paired with increasingly sophisticated models of its formation.

\acknowledgments

\added{We thank the anonymous referee, whose careful reading of this paper improved its clarity.} We also thank Emily C. Cunningham for helpful comments on the manuscript and Stephen Gwyn for reducing the photometry for slitmask f123\_1. I.E. was generously supported by a Carnegie-Princeton Fellowship through the Carnegie Observatories. This material is based upon work supported by the NSF under Grants No.\ AST-1614081 (E.N.K.), AST-1614569 (K.M.G, J.W.), and AST-1412648 (P.G.). E.N.K gratefully acknowledges support from a Cottrell Scholar award administered by the Research Corporation for Science Advancement, as well as funding from generous donors to the California Institute of Technology.

We are grateful to the many people who have worked to make the Keck Telescope and its instruments a reality and to operate and maintain the Keck Observatory. The authors wish to recognize and acknowledge the very significant cultural role and reverence that the summit of Maunakea has always had within the indigenous Hawaiian community.  We are most fortunate to have the opportunity to conduct observations from this mountain.

\vspace{5mm}
\facilities{Keck (DEIMOS)}
\software{Astropy \citep{Astropy2013, Astropy2018}, emcee \citep{Foreman-Mackey2013}} 


\bibliographystyle{aasjournal}

\begin{thebibliography}{}
\bibitem[Astropy Collaboration et al.(2013)]{Astropy2013} Astropy Collaboration, Robitaille, T.~P., Tollerud, E.~J., et al.\ 2013, \aap, 558, A33. doi:10.1051/0004-6361/201322068
\bibitem[Astropy Collaboration et al.(2018)]{Astropy2018} Astropy Collaboration, Price-Whelan, A.~M., Sip{\H{o}}cz, B.~M., et al.\ 2018, \aj, 156, 123. doi:10.3847/1538-3881/aabc4f
\bibitem[Abazajian et al.(2009)]{sdssdr72009} Abazajian, K.~N., Adelman-McCarthy, J.~K., Ag{\"u}eros, M.~A., et al.\ 2009, \apjs, 182, 543. doi:10.1088/0067-0049/182/2/543
\bibitem[Barker et al.(2007)]{Barker2007} Barker, M.~K., Sarajedini, A., Geisler, D., et al.\ 2007, \aj, 133, 1125. doi:10.1086/511185
\bibitem[Bernard et al.(2015)]{Bernard2015} Bernard, E.~J., Ferguson, A.~M.~N., Chapman, S.~C., et al.\ 2015, \mnras, 453, L113. doi:10.1093/mnrasl/slv116
\bibitem[Brown et al.(2003)]{Brown2003} Brown, T.~M., Ferguson, H.~C., Smith, E., et al.\ 2003, \apjl, 592, L17. doi:10.1086/376935
\bibitem[Brown et al.(2006)]{Brown2006} Brown, T.~M., Smith, E., Guhathakurta, P., et al.\ 2006, \apjl, 636, L89. doi:10.1086/500089
\bibitem[Bullock \& Johnston(2005)]{BullockJohnston2005} Bullock, J.~S. \& Johnston, K.~V.\ 2005, \apj, 635, 931. doi:10.1086/497422
\bibitem[Carrera et al.(2008a)]{Carrera2008smc} Carrera, R., Gallart, C., Aparicio, A., et al.\ 2008, \aj, 136, 1039. doi:10.1088/0004-6256/136/3/1039 -- commenting out the Carrera papers because they don't quantify the gradients
\bibitem[Carrera et al.(2008b)]{Carrera2008lmc} Carrera, R., Gallart, C., Hardy, E., et al.\ 2008, \aj, 135, 836. doi:10.1088/0004-6256/135/3/836
\added{\bibitem[Cheng et al.(2012)]{Cheng2012a} Cheng, J.~Y., Rockosi, C.~M., Morrison, H.~L., et al.\ 2012, \apj, 746, 149. doi:10.1088/0004-637X/746/2/149}
\bibitem[Chou et al.(2007)]{Chou2007} Chou, M.-Y., Majewski, S.~R., Cunha, K., et al.\ 2007, \apj, 670, 346. doi:10.1086/522483
\bibitem[Choudhury et al.(2016)]{Choudhury2016} Choudhury, S., Subramaniam, A., \& Cole, A.~A.\ 2016, \mnras, 455, 1855. doi:10.1093/mnras/stv2414
\bibitem[Choudhury et al.(2018)]{Choudhury2018} Choudhury, S., Subramaniam, A., Cole, A.~A., et al.\ 2018, \mnras, 475, 4279. doi:10.1093/mnras/sty087
\bibitem[Clementini et al.(2011)]{Clementini2011} Clementini, G., Contreras Ramos, R., Federici, L., et al.\ 2011, \apj, 743, 19
\bibitem[Cohen et al.(2018)]{Cohen2018} Cohen, R.~E., Kalirai, J.~S., Gilbert, K.~M., et al.\ 2018, \aj, 156, 230. doi:10.3847/1538-3881/aae52d
\bibitem[Collins et al.(2011)]{Collins2011} Collins, M.~L.~M., Chapman, S.~C., Ibata, R.~A., et al.\ 2011, \mnras, 413, 1548
\bibitem[Conn et al.(2016)]{Conn2016} Conn, A.~R., McMonigal, B., Bate, N.~F., et al.\ 2016, \mnras, 458, 3282
\bibitem[Conroy et al.(2014)]{Conroy2014} Conroy, C., Graves, G.~J., \& van Dokkum, P.~G.\ 2014, \apj, 780, 33. doi:10.1088/0004-637X/780/1/33
\bibitem[Cooper et al.(2012)]{Cooper2012} Cooper, M.~C., Griffith, R.~L., Newman, J.~A., et al.\ 2012, \mnras, 419, 3018
\bibitem[D'Souza \& Bell(2018)]{DSouzaBell2018} D'Souza, R. \& Bell, E.~F.\ 2018, Nature Astronomy, 2, 737. doi:10.1038/s41550-018-0533-x
\added{\bibitem[Dobbie et al.(2014)]{Dobbie2014b} Dobbie, P.~D., Cole, A.~A., Subramaniam, A., et al.\ 2014, \mnras, 442, 1680. doi:10.1093/mnras/stu926}
\bibitem[Dorman et al.(2015)]{Dorman2015} Dorman, C.~E., Guhathakurta, P., Seth, A.~C., et al.\ 2015, \apj, 803, 24. doi:10.1088/0004-637X/803/1/24
\bibitem[Escala et al.(2019)]{Escala2019} Escala, I., Kirby, E.~N., Gilbert, K.~M., et al.\ 2019, \apj, 878, 42
\bibitem[Escala et al.(2020a)]{Escala2020a} Escala, I., Gilbert, K.~M., Kirby, E.~N., et al.\ 2020, \apj, 889, 177
\bibitem[Escala et al.(2020b)]{Escala2020b} Escala, I., Kirby, E.~N., Gilbert, K.~M., et al.\ 2020, \apj, 902, 51. doi:10.3847/1538-4357/abb474
\bibitem[Fardal et al.(2006)]{Fardal2006} Fardal, M.~A., Babul, A., Geehan, J.~J., et al.\ 2006, \mnras, 366, 1012
\bibitem[Fardal et al.(2007)]{Fardal2007} Fardal, M.~A., Guhathakurta, P., Babul, A., et al.\ 2007, \mnras, 380, 15
\bibitem[Fardal et al.(2008)]{Fardal2008} Fardal, M.~A., Babul, A., Guhathakurta, P., et al.\ 2008, \apjl, 682, L33. doi:10.1086/590386
\bibitem[Fardal et al.(2012)]{Fardal2012} Fardal, M.~A., Guhathakurta, P., Gilbert, K.~M., et al.\ 2012, \mnras, 423, 3134. doi:10.1111/j.1365-2966.2012.21094.x
\bibitem[Fardal et al.(2013)]{Fardal2013} Fardal, M.~A., Weinberg, M.~D., Babul, A., et al.\ 2013, \mnras, 434, 2779
\bibitem[Ferguson et al.(2002)]{Ferguson2002} Ferguson, A.~M.~N., Irwin, M.~J., Ibata, R.~A., et al.\ 2002, \aj, 124, 1452. doi:10.1086/342019
\bibitem[Ferguson et al.(2005)]{Ferguson2005} Ferguson, A.~M.~N., Johnson, R.~A., Faria, D.~C., et al.\ 2005, \apjl, 622, L109. doi:10.1086/429371
\bibitem[Font et al.(2006)]{Font2006} Font, A.~S., Johnston, K.~V., Guhathakurta, P., et al.\ 2006, \aj, 131, 1436
\bibitem[Font et al.(2006)]{Font2006abund} Font, A.~S., Johnston, K.~V., Bullock, J.~S., et al.\ 2006, \apj, 638, 585. doi:10.1086/498970
\bibitem[Foreman-Mackey et al.(2013)]{Foreman-Mackey2013} Foreman-Mackey, D., Hogg, D.~W., Lang, D., et al.\ 2013, \pasp, 125, 306
\bibitem[Freeman \& Bland-Hawthorn(2002)]{FreemanBland-Hawthorn2002} Freeman, K. \& Bland-Hawthorn, J.\ 2002, \araa, 40, 487. doi:10.1146/annurev.astro.40.060401.093840
\bibitem[Gallazzi et al.(2006)]{Galazzi2006} Gallazzi, A., Charlot, S., Brinchmann, J., et al.\ 2006, \mnras, 370, 1106. doi:10.1111/j.1365-2966.2006.10548.x
\bibitem[Gallazzi et al.(2021)]{Galazzi2021} Gallazzi, A.~R., Pasquali, A., Zibetti, S., et al.\ 2021, \mnras, 502, 4457. doi:10.1093/mnras/stab265
\bibitem[Gebek \& Matthee(2021)]{Gebek2021} Gebek, A. \& Matthee, J.\ 2021, arXiv:2102.04561
\bibitem[Gilbert et al.(2006)]{Gilbert2006} Gilbert, K.~M., Guhathakurta, P., Kalirai, J.~S., et al.\ 2006, \apj, 652, 1188
\bibitem[Gilbert et al.(2007)]{Gilbert2007} Gilbert, K.~M., Fardal, M., Kalirai, J.~S., et al.\ 2007, \apj, 668, 245
\bibitem[Gilbert et al.(2009)]{Gilbert2009} Gilbert, K.~M., Guhathakurta, P., Kollipara, P., et al.\ 2009, \apj, 705, 1275
\added{\bibitem[Gilbert et al.(2014)]{Gilbert2014} Gilbert, K.~M., Kalirai, J.~S., Guhathakurta, P., et al.\ 2014, \apj, 796, 76. doi:10.1088/0004-637X/796/2/76}
\bibitem[Gilbert et al.(2018)]{Gilbert2018} Gilbert, K.~M., Tollerud, E., Beaton, R.~L., et al.\ 2018, \apj, 852, 128
\bibitem[Gilbert et al.(2019)]{Gilbert2019} Gilbert, K.~M., Kirby, E.~N., Escala, I., et al.\ 2019, \apj, 883, 128
\bibitem[Gilbert et al.(2020)]{Gilbert2020} Gilbert, K.~M., Wojno, J., Kirby, E.~N., et al.\ 2020, \aj, 160, 41. doi:10.3847/1538-3881/ab9602
\bibitem[Guhathakurta et al.(2006)]{Guhathakurta2006} Guhathakurta, P., Rich, R.~M., Reitzel, D.~B., et al.\ 2006, \aj, 131, 2497
\bibitem[Kalirai et al.(2006)]{Kalirai2006d} Kalirai, J.~S., Guhathakurta, P., Gilbert, K.~M., et al.\ 2006, \apj, 641, 268
\bibitem[Lee et al.(2015)]{Lee2015} Lee, D.~M., Johnston, K.~V., Sen, B., et al.\ 2015, \apj, 802, 48. doi:10.1088/0004-637X/802/1/48
\bibitem[Hammer et al.(2010)]{Hammer2010} Hammer, F., Yang, Y.~B., Wang, J.~L., et al.\ 2010, \apj, 725, 542. doi:10.1088/0004-637X/725/1/542
\bibitem[Hammer et al.(2018)]{Hammer2018} Hammer, F., Yang, Y.~B., Wang, J.~L., et al.\ 2018, \mnras, 475, 2754. doi:10.1093/mnras/stx3343
\bibitem[Hasselquist et al.(2017)]{Hasselquist2017} Hasselquist, S., Shetrone, M., Smith, V., et al.\ 2017, \apj, 845, 162. doi:10.3847/1538-4357/aa7ddc
\added{\bibitem[Hayden et al.(2014)]{Hayden2014} Hayden, M.~R., Holtzman, J.~A., Bovy, J., et al.\ 2014, \aj, 147, 116. doi:10.1088/0004-6256/147/5/116}
\bibitem[Hayes et al.(2020)]{Hayes2020} Hayes, C.~R., Majewski, S.~R., Hasselquist, S., et al.\ 2020, \apj, 889, 63. doi:10.3847/1538-4357/ab62ad
\bibitem[Haywood et al.(2018)]{Haywood2018} Haywood, M., Di Matteo, P., Lehnert, M.~D., et al.\ 2018, \apj, 863, 113. doi:10.3847/1538-4357/aad235
\bibitem[Helmi et al.(2018)]{Helmi2018} Helmi, A., Babusiaux, C., Koppelman, H.~H., et al.\ 2018, \nat, 563, 85. doi:10.1038/s41586-018-0625-x
\bibitem[Helmi(2020)]{Helmi2020} Helmi, A.\ 2020, \araa, 58, 205. doi:10.1146/annurev-astro-032620-021917
\added{\bibitem[Ho et al.(2015)]{Ho2015} Ho, N., Geha, M., Tollerud, E.~J., et al.\ 2015, \apj, 798, 77. doi:10.1088/0004-637X/798/2/77}
\bibitem[Hogg et al.(2010)]{Hogg2010} Hogg, D.~W., Bovy, J., \& Lang, D.\ 2010, arXiv:1008.4686
\bibitem[Hopkins et al.(2009)]{Hopkins2009} Hopkins, P.~F., Cox, T.~J., Younger, J.~D., et al.\ 2009, \apj, 691, 1168. doi:10.1088/0004-637X/691/2/1168
\bibitem[Ibata et al.(2001b)]{Ibata2001b} Ibata, R., Irwin, M., Lewis, G.~F., et al.\ 2001, \apjl, 547, L133. doi:10.1086/318894
\bibitem[Ibata et al.(2001a)]{Ibata2001} Ibata, R., Irwin, M., Lewis, G., et al.\ 2001, \nat, 412, 49
\bibitem[Ibata et al.(2004)]{Ibata2004} Ibata, R., Chapman, S., Ferguson, A.~M.~N., et al.\ 2004, \mnras, 351, 117. doi:10.1111/j.1365-2966.2004.07759.x
\bibitem[Ibata et al.(2007)]{Ibata2007} Ibata, R., Martin, N.~F., Irwin, M., et al.\ 2007, \apj, 671, 1591. doi:10.1086/522574
\bibitem[Ibata et al.(2014)]{Ibata2014} Ibata, R.~A., Lewis, G.~F., McConnachie, A.~W., et al.\ 2014, \apj, 780, 128. doi:10.1088/0004-637X/780/2/128
\bibitem[Irwin et al.(2005)]{Irwin2005} Irwin, M.~J., Ferguson, A.~M.~N., Ibata, R.~A., et al.\ 2005, \apjl, 628, L105
\bibitem[Johnston et al.(2008)]{Johnston2008} Johnston, K.~V., Bullock, J.~S., Sharma, S., et al.\ 2008, \apj, 689, 936. doi:10.1086/592228
\bibitem[Kim et al.(2002)]{Kim2002} Kim, M., Kim, E., Lee, M.~G., et al.\ 2002, \aj, 123, 244. doi:10.1086/324639
\added{\bibitem[Kacharov et al.(2017)]{Kacharov2017} Kacharov, N., Battaglia, G., Rejkuba, M., et al.\ 2017, \mnras, 466, 2006. doi:10.1093/mnras/stw3188}
\added{\bibitem[Keller et al.(2010)]{Keller2010} Keller, S.~C., Yong, D., \& Da Costa, G.~S.\ 2010, \apj, 720, 940. doi:10.1088/0004-637X/720/1/940}
\bibitem[Kirby et al.(2008)]{Kirby2008} Kirby, E.~N., Guhathakurta, P., \& Sneden, C.\ 2008, \apj, 682, 1217
\bibitem[Kirby et al.(2009)]{Kirby2009} Kirby, E.~N., Guhathakurta, P., Bolte, M., et al.\ 2009, \apj, 705, 328
\added{\bibitem[Kirby et al.(2011)]{Kirby2011} Kirby, E.~N., Lanfranchi, G.~A., Simon, J.~D., et al.\ 2011, \apj, 727, 78. doi:10.1088/0004-637X/727/2/78}
\bibitem[Kirby et al.(2013)]{Kirby2013} Kirby, E.~N., Cohen, J.~G., Guhathakurta, P., et al.\ 2013, \apj, 779, 102. doi:10.1088/0004-637X/779/2/102
\bibitem[Kirby et al.(2015)]{Kirby2015} Kirby, E.~N., Simon, J.~D., \& Cohen, J.~G.\ 2015, \apj, 810, 56
\added{\bibitem[Kirby et al.(2017)]{Kirby2017} Kirby, E.~N., Rizzi, L., Held, E.~V., et al.\ 2017, \apj, 834, 9. doi:10.3847/1538-4357/834/1/9}
\bibitem[Kirby et al.(2020)]{Kirby2020} Kirby, E.~N., Gilbert, K.~M., Escala, I., et al.\ 2020, \aj, 159, 46
\bibitem[Kirihara et al.(2014)]{Kirihara2014} Kirihara, T., Miki, Y., \& Mori, M.\ 2014, \pasj, 66, L10. doi:10.1093/pasj/psu124
\bibitem[Kirihara et al.(2017)]{Kirihara2017} Kirihara, T., Miki, Y., Mori, M., et al.\ 2017, \mnras, 464, 3509
\bibitem[Lapenna et al.(2012)]{Lapenna2012} Lapenna, E., Mucciarelli, A., Origlia, L., et al.\ 2012, \apj, 761, 33. doi:10.1088/0004-637X/761/1/33
\added{\bibitem[Law \& Majewski(2010)]{LawMajewski2010} Law, D.~R. \& Majewski, S.~R.\ 2010, \apj, 718, 1128. doi:10.1088/0004-637X/718/2/1128}
\added{\bibitem[Leaman et al.(2013)]{Leaman2013} Leaman, R., Venn, K.~A., Brooks, A.~M., et al.\ 2013, \apj, 767, 131. doi:10.1088/0004-637X/767/2/131}
\bibitem[Lehner et al.(2020)]{Lehner2020} Lehner, N., Berek, S.~C., Howk, J.~C., et al.\ 2020, \apj, 900, 9. doi:10.3847/1538-4357/aba49c
\bibitem[Mackereth et al.(2018)]{Mackereth2018} Mackereth, J.~T., Crain, R.~A., Schiavon, R.~P., et al.\ 2018, \mnras, 477, 5072. doi:10.1093/mnras/sty972
\bibitem[Marigo et al.(2017)]{Marigo2017} Marigo, P., Girardi, L., Bressan, A., et al.\ 2017, \apj, 835, 77
\bibitem[Martin et al.(2013)]{Martin2013} Martin, N.~F., Ibata, R.~A., McConnachie, A.~W., et al.\ 2013, \apj, 776, 80. doi:10.1088/0004-637X/776/2/80
\bibitem[McConnachie et al.(2003)]{McConnachie2003} McConnachie, A.~W., Irwin, M.~J., Ibata, R.~A., et al.\ 2003, \mnras, 343, 1335
\bibitem[McConnachie et al.(2005)]{McConnachie2005} McConnachie, A.~W., Irwin, M.~J., Ferguson, A.~M.~N., et al.\ 2005, \mnras, 356, 979. doi:10.1111/j.1365-2966.2004.08514.x
\bibitem[McConnachie et al.(2018)]{McConnachie2018} McConnachie, A.~W., Ibata, R., Martin, N., et al.\ 2018
\added{\bibitem[Mercado et al.(2021)]{Mercado2021} Mercado, F.~J., Bullock, J.~S., Boylan-Kolchin, M., et al.\ 2021, \mnras, 501, 5121. doi:10.1093/mnras/staa3958}
\bibitem[Miki et al.(2016)]{Miki2016} Miki, Y., Mori, M., \& Rich, R.~M.\ 2016, \apj, 827, 82. doi:10.3847/0004-637X/827/1/82
\bibitem[Monaco et al.(2007)]{Monaco2007} Monaco, L., Bellazzini, M., Bonifacio, P., et al.\ 2007, \aap, 464, 201. doi:10.1051/0004-6361:20066228
\bibitem[Mori \& Rich(2008)]{MoriRich2008} Mori, M. \& Rich, R.~M.\ 2008, \apjl, 674, L77. doi:10.1086/529140
\bibitem[Mucciarelli et al.(2017)]{Mucciarelli2017} Mucciarelli, A., Bellazzini, M., Ibata, R., et al.\ 2017, \aap, 605, A46. doi:10.1051/0004-6361/201730707
\bibitem[Naidu et al.(2020)]{Naidu2020} Naidu, R.~P., Conroy, C., Bonaca, A., et al.\ 2020, \apj, 901, 48. doi:10.3847/1538-4357/abaef4
\bibitem[Naiman et al.(2018)]{Naiman2018} Naiman, J.~P., Pillepich, A., Springel, V., et al.\ 2018, \mnras, 477, 1206. doi:10.1093/mnras/sty618
\bibitem[Newman et al.(2013)]{Newman2013} Newman, J.~A., Cooper, M.~C., Davis, M., et al.\ 2013, \apjs, 208, 5
\bibitem[Nidever et al.(2020)]{Nidever2020} Nidever, D.~L., Hasselquist, S., Hayes, C.~R., et al.\ 2020, \apj, 895, 88. doi:10.3847/1538-4357/ab7305
\added{\bibitem[Parisi et al.(2016)]{Parisi2016} Parisi, M.~C., Geisler, D., Carraro, G., et al.\ 2016, \aj, 152, 58. doi:10.3847/0004-6256/152/3/58}
\bibitem[Pomp{\'e}ia et al.(2008)]{Pompeia2008} Pomp{\'e}ia, L., Hill, V., Spite, M., et al.\ 2008, \aap, 480, 379. doi:10.1051/0004-6361:20064854
\bibitem[Richardson et al.(2008)]{Richardson2008} Richardson, J.~C., Ferguson, A.~M.~N., Johnson, R.~A., et al.\ 2008, \aj, 135, 1998. doi:10.1088/0004-6256/135/6/1998
\bibitem[Robertson et al.(2005)]{Robertson2005} Robertson, B., Bullock, J.~S., Font, A.~S., et al.\ 2005, \apj, 632, 872. doi:10.1086/452619
\bibitem[Sadoun et al.(2014)]{Sadoun2014} Sadoun, R., Mohayaee, R., \& Colin, J.\ 2014, \mnras, 442, 160. doi:10.1093/mnras/stu850
\bibitem[Segers et al.(2016)]{Segers2016} Segers, M.~C., Schaye, J., Bower, R.~G., et al.\ 2016, \mnras, 461, L102. doi:10.1093/mnrasl/slw111
\added{\bibitem[Shetrone et al.(2001)]{Shetrone2001} Shetrone, M.~D., C{\^o}t{\'e}, P., \& Sargent, W.~L.~W.\ 2001, \apj, 548, 592. doi:10.1086/319022}
\bibitem[Simon \& Geha(2007)]{SimonGeha2007} Simon, J.~D. \& Geha, M.\ 2007, \apj, 670, 313
\added{\bibitem[VandenBerg et al.(2006)]{VandenBerg2006} VandenBerg, D.~A., Bergbusch, P.~A., \& Dowler, P.~D.\ 2006, \apjs, 162, 375. doi:10.1086/498451}
\bibitem[Tanaka et al.(2010)]{Tanaka2010} Tanaka, M., Chiba, M., Komiyama, Y., et al.\ 2010, \apj, 708, 1168. doi:10.1088/0004-637X/708/2/1168
\bibitem[Tiede et al.(2004)]{Tiede2004} Tiede, G.~P., Sarajedini, A., \& Barker, M.~K.\ 2004, \aj, 128, 224. doi:10.1086/421369
\bibitem[Thomas et al.(2005)]{Thomas2005} Thomas, D., Maraston, C., Bender, R., et al.\ 2005, \apj, 621, 673. doi:10.1086/426932
\bibitem[Van der Swaelmen et al.(2013)]{VanderSwaelmen2013} Van der Swaelmen, M., Hill, V., Primas, F., et al.\ 2013, \aap, 560, A44. doi:10.1051/0004-6361/201321109
\added{\bibitem[Vargas et al.(2014)]{Vargas2014a} Vargas, L.~C., Geha, M.~C., \& Tollerud, E.~J.\ 2014, \apj, 790, 73. doi:10.1088/0004-637X/790/1/73}
\bibitem[Vargas et al.(2014)]{Vargas2014b} Vargas, L.~C., Gilbert, K.~M., Geha, M., et al.\ 2014, \apjl, 797, L2
\added{\bibitem[Venn et al.(2004)]{Venn2004} Venn, K.~A., Irwin, M., Shetrone, M.~D., et al.\ 2004, \aj, 128, 1177. doi:10.1086/422734}
\bibitem[Williams et al.(2015)]{Williams2015} Williams, B.~F., Dalcanton, J.~J., Dolphin, A.~E., et al.\ 2015, \apj, 806, 48. doi:10.1088/0004-637X/806/1/48
\bibitem[Wojno et al.(2020)]{Wojno2020} Wojno, J., Gilbert, K.~M., Kirby, E.~N., et al.\ 2020, \apj, 895, 78
\end{thebibliography}

\end{document}